\documentclass[11pt]{article}
\usepackage{amsmath,amssymb,color}

\usepackage{amsfonts}

\makeatletter \@addtoreset{equation}{section} \makeatother

\newcommand{\hoch}[1]{$\, ^{#1}$}

\newcommand{\be}{\begin{equation}}
\newcommand{\ee}{\end{equation}}
\newcommand{\bea}{\setlength\arraycolsep{2pt} \begin{eqnarray}}
\newcommand{\eea}{\end{eqnarray}}

\def\ft#1#2{{\textstyle{\frac{\scriptstyle #1}{\scriptstyle #2} } }}
\def\fft#1#2{{\frac{#1}{#2}}}

\def\0{{\sst{(0)}}}
\def\1{{\sst{(1)}}}
\def\2{{\sst{(2)}}}
\def\3{{\sst{(3)}}}
\def\4{{\sst{(4)}}}
\def\5{{\sst{(5)}}}
\def\6{{\sst{(6)}}}
\def\7{{\sst{(7)}}}
\def\8{{\sst{(8)}}}
\def\sst#1{{\scriptscriptstyle #1}}
\def\oneone{\rlap 1\mkern4mu{\rm l}}

\def\del{{\partial}}

\begin{document}

\begin{flushright}
\hfill{ \
MIFPA-11-06\ \ \ \ KIAS-P11006 }
\end{flushright}

\vspace{25pt}
\begin{center}
{\large {\bf Pseudo-Killing Spinors, Pseudo-supersymmetric
$p$-branes,\\ Bubbling and Less-bubbling AdS Spaces}}

\vspace{15pt}

H. L\"u\hoch{1,2} and Zhao-Long Wang\hoch{3}

\vspace{10pt}

\hoch{1}{\it China Economics and Management Academy\\
Central University of Finance and Economics, Beijing 100081}

\vspace{10pt}

\hoch{2}{\it Institute for Advanced Study, Shenzhen
University, Nanhai Ave 3688, Shenzhen 518060}

\vspace{10pt}

\hoch{3} {\it School of Physics, Korea Institute for Advanced Study,
Seoul 130-722, Korea}

\vspace{40pt}

\underline{ABSTRACT}
\end{center}

We consider Einstein gravity coupled to an $n$-form field strength
in $D$ dimensions.  Such a theory cannot be supersymmetrized in
general, we nevertheless propose a pseudo-Killing spinor equation
and show that the AdS$\times$Sphere vacua have the maximum number of
pseudo-Killing spinors, and hence are fully pseudo-supersymmetric.
We show that extremal $p$-branes and their intersecting
configurations preserve fractions of the pseudo-supersymmetry.  We
study the integrability condition for general $(D,n)$ and obtain the
additional constraints that are required so that the existence of
the pseudo-Killing spinors implies the Einstein equations of motion.
We obtain new pseudo-supersymmetric bubbling AdS$_5\times S^5$
spaces that are supported by a non-self-dual 5-form.  This
demonstrates that non-supersymmegtric conformal field theories may
also have bubbling states of arbitrary droplets of free fermions in
the phase space. We also obtain an example of less-bubbling AdS
geometry in $D=8$, whose bubbling effects are severely restricted by
the additional constraint arising from the integrability condition.

\vspace{15pt}

\thispagestyle{empty}

\pagebreak
\setcounter{page}{1}

\tableofcontents

\addtocontents{toc}{\protect\setcounter{tocdepth}{2}}


\newpage

\section{Introduction}

The original proposal of the AdS/CFT correspondence
\cite{mald,gkp,wit} is under the framework of supergravity or
superstring. Supersymmetry provides a powerful organization to test
the correspondence.  The holographic principle underlying the
AdS/CFT correspondence, however, goes beyond supersymmetry.
Application of the AdS/CFT correspondence in non-supersymmetric
gauge theories has been a flourishing research topic since the
inception of the AdS/CFT correspondence.

The AdS/CFT correspondence connects the strongly-coupled gauge
theory in the boundary of the anti-de Sitter spacetime (AdS) with
classical gravity in the bulk. The key to apply the correspondence
is to construct the appropriate classical gravity background that is
dual to the boundary field theory which is typically impossible to
study on its own due to its strong-coupling nature.  The
construction of the bulk geometry belongs to the conventional
subject of general relativity. There is a significant advantage of
supersymmetry which enables one to construct a large class of BPS
solutions that preserve a certain fraction of the maximum number of
the supersymmetry of the theory.  For simple supergravities, it was
demonstrated that the assumption of the existence of a Killing
spinor, a defining property of BPS solutions, enables one to derive
all the supersymmetric solutions of the theory \cite{gghpr}.
Employing the same technique, the most general $\fft12$-BPS
solutions in type IIB supergravity supported by the self-dual 5-form
field strength was constructed in \cite{llm}.  These solutions,
called LLM bubbling AdS spaces, describe smooth geometries
asymptotic to AdS$_5\times S^5$ and can be reduced to solutions of
certain linear Laplace equation with two-dimensional bubble-like
boundary conditions.  These smooth geometries are dual to arbitrary
droplets of free fermions in phase space of the dual conformal field
theory.  Such solutions are very unlikely to be found without the
aid of supersymmetry.

One motivation of this paper is to investigate whether it is
possible to construct such highly non-trivial bubbling geometry in a
theory that cannot be supersymmetrized.  This will resolve the issue
whether arbitrary free-fermion droplets can also arise in a
non-supersymmetric gauge theory. The bubbling AdS spaces obtained in
type IIB supergravity and M-theory involve very simple field
configurations: gravity coupled with the self-dual 5-form or 4-form
field strengths respectively.  Indeed such a system provides the
simplest origin of the cosmological constant in lower dimensions. We
thus consider Einstein gravity coupled to an $n$-form field strength
in general $D$ dimensions.  For specific cases, namely
$(D,n)=(11,4), (10,5), (6,3), (5,2), (4,2)$, the system becomes
(part of) the bosonic action of a supersymmetric theory. (To be
precise, in 10 and 6 dimensions, supersymmetry requires that the
5-form and 3-form be self-dual.) In general, however, the $(D,n)$
system cannot be made to or part of a supersymmetry theory.  Our
goal is to investigate whether such an intrinsically
non-supersymmetric system may nevertheless admit solutions with
characteristics of BPS solutions in supergravities. For simple
cohomogeneity-one classes, it was known that even the
Schwarzschild-AdS black hole can be solved by the super-potential
method in which the second-order differential equations can be
successfully reduced to a set of first-order equations {\it via} a
super potential \cite{clv,lpv}. (See also \cite{psrv}.) This
suggests that certain non-supersymmetric systems may exhibit
characteristics of supersymmetry; they are {\it
pseudo}-supersymmetric. Pseudo-supersymmetry for gravity coupled to
scalars were introduced in \cite{sketow}. Pseudo-supersymmetry for
de Sitter ``supergravity" was discussed in \cite{ds1,ds2}.

In section 2, we introduce a pseudo-Killing spinor equation for the
$(D,n)$ system. It involves one $(n-1)$-gamma structure and one
$(n+1)$-gamma structure,
\begin{equation}
D_M \epsilon + \fft{\tilde\alpha}{(n-1)!} \Gamma^{M_1\cdots M_{n-1}}
F_{M M_1\cdots M_{n-1}}\epsilon + \fft{\tilde\beta}{n!}
\Gamma_M{}^{M_1\cdots M_n} F_{M_1\cdots M_n}
\epsilon=0\,,\label{ksdef0}
\end{equation}
with two constant parameters $(\tilde\alpha,\tilde\beta)$ to be
determined, as follows. The $(D,n)$ system admits two
AdS$\times$Sphere vacua with the $n$-form carries either the
electric or the magnetic fluxes.  We can fix the value of $\tilde
\alpha$ and $\tilde \beta$ up to a relative sign, by requiring that
the vacua have the maximum number of allowed pseudo-Killing spinors.
These pseudo-Killing spinors are tensor products of real Killing
spinors in the AdS spaces and spheres. Note that the overall sign of
$(\tilde \alpha,\tilde \beta)$ can be absorbed into the $n$-form
field strength $F$.  We then obtain the explicit pseudo-Killing
spinors for the AdS$\times$Sphere vacua in section 3.

   In section 4, we obtained both the electric and magnetic extremal
$p$-branes for the $(D,n)$ system. We show, by explicit
construction, that the existence of the pseudo-Killing spinors for
these solutions fixes the relative sign of $(\tilde
\alpha,\tilde\beta)$. This thus fully determines the pseudo-Killing
spinor equation.  It turns out that the $p$-branes all preserve half
of the maximum number of the pseudo-Killing spinors of the vacua.
It should be emphasized that in the special cases mentioned above
for which the system can be indeed supersymmetrized, the
pseudo-Killing spinors become real Killing spinors of the
supersymmetric theory. The corresponding brane solutions are the
previously-known BPS $p$-branes in supergravities. In section 5, we
obtain a large class of pair-wise intersecting $p$-branes, and
construct the corresponding pseudo-Killing spinors.

  Even in a supersymmetric theory, the existence of Killing spinor of
a bosonic configuration does not always imply that it would satisfy
the equations of motion. Additional constraints have to be imposed,
{\it e.g.} the Killing vector constructed from the Killing spinor
has to be time-like or null. In section 6, we investigate the
integrability condition of the pseudo-Killing spinor equation for
generic $(D,n)$.  We obtain additional algebraic quadratic
constraints on the form fields.  These constraints, although
satisfied by the vacua and $p$-branes discussed earlier, give rise
to severe restrictions on possible solutions one may have.  For
$n=2,3,4,5$, there exist critical dimensions in each case,
corresponding precisely to relevant supergravities, where such a
restriction vanishes.

  For $n=5$, the critical dimension is 10 and all the constraints
vanish if the 5-form is self-dual.  The resulting theory is then the
$SL(2,R)$ singlet of type IIB supergravity.  If we relax the
requirement that the 5-form be self-dual, there are extra
constraints from the integrability condition. However, we find that
these conditions can be satisfied provided that the non-vanishing
components of the 5-form are restricted to lie in a sub-manifold of
seven dimensions.  This property enables us to construct explicitly
a new pseudo-supersymmetric bubbling geometry in the $(D,n)=(10,5)$
system. The result is presented in section 7.  The 5-form is not
self-dual and cannot be made so by adjusting parameters in the
solution. Thus the system is intrinsically non-supersymmetric and
cannot be embedded in type IIB supergravity. Our construction
demonstrates that bubbling AdS geometries are not uniquely possessed
by supergravities, and consequently, some strongly-coupled
non-supersymmetric conformal gauge theory in four-dimensions can
also have bubbling states, {\it i.e.} the arbitrary droplets of free
fermions in the phase space.

  In section 8, we construct analogous configurations in $(D,n)=(8,4)$
to examine the effect of additional constraints on the bubbling
nature of the asymptotic AdS$_4\times S^4$ geometries.  Although the
pseudo-Killing spinor equation still leads to a linear Laplace
equation, an additional non-linear constraint has to be imposed from
the integrability condition.  This implies that linear supposition
of the solutions, the key to the bubbling effect, is no longer
applicable, and the geometry becomes less bubbling. Although we are
unable to obtain the most general solution for the non-linear
system, we argue that it is not unreasonable to expect that
additional less-bubbling solutions beyond the vacua may still exist,
which may reveal some specific droplets of free fermions that are
allowed in the phase space.

  We conclude our paper in section 9.  Appendices A and B are detail
construction of the bubbling and less-bubbling geometries presented
in sections 7 and 8 respectively.

\section{The theory and pseudo-Killing spinors}
\label{thetheory}

The theory we consider throughout this paper is the Einstein-Hilbert
action coupled to an $n$-form field strength. The Lagrangian is
given by
\begin{equation}
{\cal L} = \sqrt{-g} \Big(R - \fft{1}{2\, n!}
F_{(n)}^2\Big)\,,\label{genlag}
\end{equation}
where $F_{(n)}=dA_{(n-1)}$.  The Bianchi identity and the equation
of motion for the $n$-form are
\begin{eqnarray}
dF_{(n)}=0 &\longrightarrow& \partial_{[M} F_{M_1M_2\cdots
M_n]}=0\,,\cr 
d{*F_{(n)}}=0 &\longrightarrow& \fft{1}{\sqrt{-g}} \partial_{M_1}
F^{M_1M_2\cdots M_n}=0\,.\label{formeom} 
\end{eqnarray}
The Einstein equations of motion are given by
\begin{equation}
R_{MN}=\fft{1}{2\, (n-1)!} \Big(F_{MN}^2 - \fft{n-1}{n(D-2)} F^2
g_{MN}\Big)\,.\label{einsteineom}
\end{equation}
The system admits AdS$_n\times S^{D-n}$ and AdS$_{D-n}\times S^n$
vacuum solutions, carrying electric and magnetic fluxes
respectively.  The AdS$_n\times S^{D-n}$ solution is given by
\begin{equation}
ds_D^2 = ds_{{\rm AdS}_n}^2 + d\Sigma_{D-n}^2\,,\qquad F_{(n)} =
\fft{2\lambda d}{\sqrt{\Delta}} \omega_{{\rm
AdS}_n}\,,\label{adssphere1}
\end{equation}
where $ds_{{\rm AdS}_n}^2$ and $d\Sigma_{D-n}^2$ are metrics for
AdS$_n$ and $S^{D-n}$ respectively, with
\begin{equation}
R_{\mu\nu}=-(n-1) \lambda^2 g_{\mu\nu}\,,\qquad R_{ij}=\tilde d
\tilde \lambda^2 g_{ij}\,,\qquad \lambda d= \tilde \lambda \tilde
d\,,
\end{equation}
and $\omega_{{\rm AdS}_n}$ is the volume form for $ds_{{\rm
AdS}_n}^2$.  Note that for convenience, we have introduced
\begin{equation}
d=n-1\,,\qquad \tilde d=D-n-1\,,\qquad \Delta = \fft{2d\tilde
d}{D-2}\,.
\end{equation}
The AdS$_{D-n}\times S^n$ solution is given by
\begin{equation}
ds_D^2=ds_{AdS_{D-n}}^2 + d\Sigma_n^2\,,\qquad F_{(n)} =
\fft{2\tilde\lambda\tilde d}{\sqrt{\Delta}} \omega_{S^n}\,,
\label{adssphere2}
\end{equation}
where
\begin{equation}
R_{\mu\nu}=-\tilde d\tilde\lambda^2 g_{\mu\nu}\,,\qquad
R_{ij}=d\lambda^2 g_{ij}\,,\qquad \lambda d = \tilde \lambda \tilde
d\,.
\end{equation}

    We now introduce a spinor $\hat \epsilon$ that satisfies the
following equation
\begin{equation}
D_M \hat \epsilon + \fft{\tilde\alpha}{(n-1)!} \Gamma^{M_1\cdots M_{n-1}}
F_{M M_1\cdots M_{n-1}}\hat \epsilon + \fft{\tilde\beta}{n!}
\Gamma_M{}^{M_1\cdots M_n} F_{M_1\cdots M_n}\hat
\epsilon=0\,,\label{ksdef1}
\end{equation}
where $D_M$ is covariant derivative defined by
\begin{equation}
D_M\hat \epsilon \equiv \partial_M \hat \epsilon+ \ft14
(\omega_{M})^a{}_b \Gamma_a{}^b\hat \epsilon\,.
\end{equation}
The constant parameters $(\tilde \alpha, \tilde \beta)$ are to be
determined. For vanishing $F$, the equation (\ref{ksdef1}) defines
the standard Killing spinors in Ricci-flat backgrounds.  Just as the
standard Ricci-flat Killing spinors which may exist in a
non-supersymmetric theory, the definition of our generalized Killing
spinor does not have to depend on whether the theory (\ref{genlag})
can be supersymmetrized or not. However, we wish that there exist
solutions of (\ref{genlag}) that admit the generalized Killing
spinors.  In particular, we impose a condition that (\ref{ksdef1})
gives rise to the maximum number of generalized Killing spinors for
both the AdS$\times$Sphere vacua, which are tensor products of real
Killing spinors in the AdS and sphere spaces.  In section 3, we show
by explicit construction, that existence of maximum number of
allowed such spinors in the AdS$\times$Sphere vacua enables us to
fix the parameters $\tilde\alpha$ and $\tilde\beta$, given by
\begin{equation}
\tilde\alpha={\rm i}^{[(n+1)/2]}\, \fft{ \sqrt{\Delta}}{4d}\,,\qquad
d\tilde\alpha + \tilde d\tilde\beta =0\,,\label{ksdef11}
\end{equation}
To be precise, the parameters $(\tilde\alpha,\tilde\beta)$ can be
fixed by this consideration up to a relative sign.  (The overall
sign of $(\tilde\alpha,\tilde\beta)$ can be absorbed in to $F$.) The
given relative sign in the above equation is actually fixed later in
section 4, by requiring that extremal $p$-branes also admit the
generalized Killing spinors.

Since the theory (\ref{genlag}) are generically non-supersymmetric,
we shall call the spinors satisfying (\ref{ksdef1}) pseudo-Killing
spinors.  In special cases such as $(D,n)=$(11,4), (10,5), (6,3),
(5,2), (4,2), the pseudo-Killing spinors become real ones. (Appropriate
chirality conditions must be imposed in even dimensions.)  In this
paper, we construct solutions of the theory (\ref{genlag}) with
pseudo-Killing spinors. We call these solutions pseudo-supersymmetric.

\section{Pseudo-Killing spinors in AdS$\times$Sphere}

In section \ref{thetheory}, we introduce the concept of
pseudo-Killing spinors for the theory (\ref{genlag}) in arbitrary
dimensions. We remarked that the parameters $\tilde\alpha$ and
$\tilde\beta$ in (\ref{ksdef1}) are fixed, up to a relative sign, so
that the AdS$\times$Sphere solutions (\ref{adssphere1}) and
(\ref{adssphere2}) have the maximum number of pseudo-Killing
spinors. In this section, we demonstrate this by obtaining the
pseudo-Killing spinors explicitly.  The analysis for pseudo-Killing
spinors in AdS$\times$Sphere vacua resembles that for real Killing
spinors in such vacua in supergravities \cite{lpr}. The full
spacetime indices $(M, N, \cdots)$ are now split into
$(\mu,\nu,\cdots)$ and $(i,j,\cdots)$ which are indices for the AdS
and sphere respectively. Since the decomposition of the
$D$-dimensional gamma matrices into those of $n$ and $(D-n)$
dimensions depends on whether $n$ and $(D-n)$ are odd or even
numbers, there are four cases \cite{lpr} to consider.

\bigskip
\noindent{\bf Case 1: $(n,D-n)=$(even, odd)}
\medskip

    In this case, the gamma matrices can be decomposed as follows
\begin{equation}
\hat \Gamma_\mu = \Gamma_\mu \otimes \oneone\,,\qquad \hat \Gamma_i
=\gamma\otimes \Gamma_i\,,
\end{equation}
where $\gamma$ is the chirality operator formed from the product of
the $\Gamma_\mu$ matrices, satisfying $\gamma^2=1$.  Thus we may
have
\begin{equation}
\gamma ={\rm i}^{(n-2)/2} \Gamma^{01\cdots(n-1)}\,.
\end{equation}
We find that the pseudo-Killing spinor in the AdS$_n\times S^{D-n}$
vacuum is given by
\begin{equation}
\hat \epsilon=\epsilon\otimes \eta\,,\label{vacuumksform}
\end{equation}
where $\epsilon$ and $\eta$ are real Killing spinors in the AdS$_n$ and
$S^{D-n}$ respectively, satisfying
\begin{eqnarray}
\hbox{AdS}_n:&& D_\mu \epsilon = \fft{{\rm i}\,\lambda}{2} \gamma
\Gamma_\mu \epsilon\,,\cr 
S^{D-n}:&& D_i\eta = \pm \fft{{\rm i}\,\tilde\lambda}{2}
\Gamma_i\eta\,.
\end{eqnarray}
Explicit construction of Killing spinors in AdS space-times and
spheres can be found in \cite{lpt,lpr}.

      For the AdS$_{D-n}\times S^n$ vacuum, the gamma matrix
decomposition is given by
\begin{equation}
\hat \Gamma_\mu = \Gamma_\mu \otimes \gamma\,,\qquad \hat \Gamma_i
=\oneone\otimes \Gamma_i\,,
\end{equation}
with
\begin{equation}
\gamma = {\rm i}^{n/2} \Gamma^{\tilde 1\tilde 2\cdots \tilde n}\,.
\end{equation}
The pseudo-Killing spinor again takes the form (\ref{vacuumksform}),
but with
\begin{eqnarray}
\hbox{AdS}_{D-n}:&& D_\mu \epsilon = \fft{\tilde\lambda}{2}
\Gamma_\mu \epsilon\,,\cr 
S^n:&& D_i\eta = \fft{\lambda}{2}\gamma \Gamma_i\eta\,.
\end{eqnarray}

\bigskip
\noindent{\bf Case 2: $(n,D-n)=$(odd, even)}
\medskip

The exercise is analogous to the previous one, and we shall present
just the results
\begin{eqnarray}
\hbox{AdS}_n\times S^{D-n}:&& \hat \Gamma_\mu = \Gamma_\mu \otimes
\gamma\,,\quad \hat \Gamma_i =\oneone\otimes \Gamma_i\,,\cr 
&&\gamma={\rm i}^{(D-n)/2} \Gamma^{\tilde 1\tilde 2\cdots
{(D-n)}}\,,\qquad {\rm i}^{[(n+1)/2]} \Gamma^{01\cdots
(n-1)}=-1\,,\cr 
&& D_\mu \epsilon = \fft{\lambda}{2}\Gamma_\mu \epsilon\,,\qquad
D_i\eta= \fft{\tilde \lambda}{2}\gamma \Gamma_i \eta\,;\cr 
\hbox{AdS}_{D-n}\times S^{n}:&& \hat \Gamma_\mu = \Gamma_\mu \otimes
\oneone\,,\quad \hat \Gamma_i =\gamma\otimes \Gamma_i\,,\cr
&& \gamma ={\rm i}^{(D-n)/2} \Gamma^{\tilde 1\tilde 2\cdots
{(D-n)}}\,,\qquad {\rm i}^{[(n+1)/2]} \Gamma^{\tilde 1\tilde 2\cdots\tilde
n}=-{\rm i}\,,\cr 
&& D_\mu \epsilon = \fft{{\rm i}\,\tilde\lambda}{2}\gamma\Gamma_\mu
\epsilon\,,\qquad D_i\eta= \fft{{\rm i}\,\lambda}{2}\Gamma_i \eta\,.
\end{eqnarray}

\bigskip
\noindent{\bf Case 3: $(n,D-n)=$(even, even)}
\medskip

There are two ways to decompose the gamma matrices.  The first is
given by
\begin{equation}
\hat \Gamma_\mu = \Gamma_\mu \otimes \oneone\,,\qquad \hat \Gamma_i
=\gamma\otimes \Gamma_i\,.
\end{equation}
In this case, the pseudo-Killing spinors for the electric
AdS$\times$Sphere metrics take the same form as those given in case
1; those for the magnetic solutions take the same form as those
given in case 2.

    Alternatively, we can decompose the gamma matrix as follows
\begin{equation}
\hat \Gamma_\mu = \Gamma_\mu \otimes \gamma\,,\qquad \hat \Gamma_i
=\oneone\otimes \Gamma_i\,.
\end{equation}
The pseudo-Killing spinors then take the same form as those given in
case 2 for the electric solutions and case 1 for the magnetic
solutions.

\bigskip
\noindent{\bf Case 4: $(n,D-n)=$(odd, odd)}
\medskip

In this case, the gamma matrices are decomposed as follows
\begin{equation}
\hat \Gamma_{\mu} = \sigma_1\otimes \Gamma_\mu\otimes
\oneone\,,\qquad \hat \Gamma_{i} = \sigma_2\otimes \oneone \otimes
\Gamma_i\,,
\end{equation}
where $\sigma_1$ and $\sigma_2$ are Pauli matrices.  For the
electric AdS$_n\times S^{(D-n)}$ solution, the pseudo-Killing spinors
take the form as
\begin{equation}
\hat\epsilon=\oneone \otimes \epsilon \otimes \eta\,,
\end{equation}
where
\begin{equation}
D_\mu \epsilon =\fft{\lambda}2 \Gamma_\mu \epsilon\,,\qquad D_i\eta=
\fft{{\rm i}\, \tilde \lambda }{2} \Gamma_i \eta\,.
\end{equation}
For the magnetic AdS$_{D-n}\times S^n$ solution, the pseudo-Killing
spinors take the same form, but with $\lambda$ and $\tilde \lambda$
switched.

Thus we have demonstrated that the AdS$\times$Sphere vacuum
solutions of (\ref{genlag}) admit the maximum number of the
pseudo-Killing spinors, defined by (\ref{ksdef1}). It should be
remarked however that the existence of the pseudo-Killing spinors
for these solutions does not fix the relative sign of the parameters
$\tilde\alpha$ and $\tilde \beta$.  The relation $d\tilde\alpha
-\tilde d\tilde \beta=0$, instead of the one given in
(\ref{ksdef11}), works equally well, provided that the orientations
of the Killing spinors in AdS and spheres are adjusted
appropriately.  However, the relative sign choice in (\ref{ksdef11})
can be fixed by requiring that extremal $p$-brane solitons also
admit pseudo-Killing spinors.  We shall discuss this in the next
section.

\section{Pseudo-supersymmetric $p$-branes}

\subsection{Electric branes}

The Lagrangian (\ref{genlag}) admits electrically-charged
$(n-2)$-brane, for which the full spacetime $(x^M)$ is split into
the $d=n-1$ dimensional world volume with coordinates $x^\mu$ and
$(D-d)$ dimensional transverse space with coordinates $y^m$.  The
solution is given by
\begin{eqnarray}
ds^2&=&e^{2A} dx^\mu dx_\mu + e^{2B} dy^i dy^i\,,\qquad
e^{2A}=H^{-2/d}\,,\qquad e^{2B}=H^{2/\tilde d}\,,\cr 
F_{(n)} &=& \fft{2}{\sqrt{\Delta}} dH^{-1}\wedge dt\wedge
dx^1\wedge\cdots dx^{n-2}\,,\label{electricp}
\end{eqnarray}
where $H$ is a harmonic function in the flat transverse space
$ds_{D-d}^2 = dy^i dy^i$. (See, {\it e.g.} \cite{stainless}.)

     We now calculate the pseudo-Killing spinor for this bosonic
$p$-brane background. A convenient choice for the metric in
(\ref{electricp}) is given by
\begin{equation}
e^{\bar \mu} = e^{A} dx^\mu\,,\qquad e^{\bar i}= e^{B} dy^i\,.
\end{equation}
Here we use barred letters to denote the tangent flat indices.  The
non-vanishing components of the corresponding spin connection are
given by
\begin{equation}
\omega_{\bar \mu\bar i}=e^{-B}\partial_i A e^{\bar \mu}\,,\qquad
\omega_{\bar i\bar j}=e^{-B} (\partial_j B e^{\bar i} - \partial_i B
e^{-\bar j})\,.
\end{equation}
Thus the covariant derivative on the spinor $\epsilon$ is given by
\begin{equation}
D\epsilon = d\epsilon + \ft12 \partial_{\bar i} A\, e^{\bar \mu}
\Gamma_{\bar \mu \bar i}\epsilon + \ft12 \partial_{\bar i} B\,
e^{\bar j} \Gamma_{\bar j\bar i} \epsilon\,.
\end{equation}
The non-vanishing components of the field strength are given by
\begin{equation}
F_{\bar i\bar \mu_1\cdots\bar \mu_{n-1}} = - \fft{2}{\sqrt\Delta}
H^{-1} \del_{\bar i} H \epsilon_{\bar \mu_1\cdots \bar \mu_{n-1}}\,.
\end{equation}
Substituting these into the generalized Killing spinor equation
(\ref{ksdef1}), we have
\begin{eqnarray}
(\partial_{\bar i}\epsilon - \ft12\partial_{\bar i} A\Gamma\epsilon)
+ \ft12 \Gamma_{\bar i\bar j} \partial_{\bar j} B (1 - \Gamma)
\epsilon &=& 0\,,\label{pbraneks1}\\
\partial_{\bar \mu}\epsilon + \ft12 \partial_{\bar i} A\, \Gamma_{\bar \mu
\bar i} (1 - \Gamma)\epsilon &=& 0\,,\label{pbraneks2}
\end{eqnarray}
where
\begin{equation}
\Gamma \equiv \fft{{\rm i}^{[(n+1)/2]}}{(n-1)!} \epsilon_{\bar
\mu_1\cdots\bar \mu_{n-1}} \Gamma^{\bar \mu_1\cdots\bar
\mu_{n-1}}\,.
\end{equation}
Thus the generalized Killing spinor is given by
\begin{equation}
\epsilon=e^{\fft12A} \epsilon_0\,,\label{fakeksres1}
\end{equation}
where $\epsilon_0$ is a constant spinor, satisfying the projection
\begin{equation}
\Gamma \epsilon_0=\epsilon_0\,.\label{fakeksres2}
\end{equation}
Owing to this projection, the number of pseudo-Killing spinors is
half of the maximum number possessed by the vacua discussed in the
previous section.  Thus the extremal $p$-branes are half pseudo-BPS
solutions.

It should be pointed out that if we had $d\tilde\alpha - \tilde
d\tilde\beta=0$, instead of the relation given in (\ref{ksdef11}),
the $(1-\Gamma)$ factor in (\ref{pbraneks1}) would become $(1 +
\Gamma)$, and hence there could be no pseudo-Killing spinors.  Thus
the existence of pseudo-Killing spinors for the extremal $p$-branes
determines the relative sign choice between $\tilde\alpha$ and
$\tilde\beta$.

\subsection{Magnetic branes}

The Lagrangian (\ref{genlag}) also admits magnetically-charged
$(D-n-2)$-brane, for which the full spacetime $(x^M)$ is split into
the $\tilde d=D-n-1$ dimensional world volume with coordinates
$x^\mu$ and $(D-\tilde d)$ dimensional transverse space with
coordinates $y^m$. The solution is given by
\begin{eqnarray}
ds^2&=&e^{2A} dx^\mu dx_\mu + e^{2B} dy^i dy^i\,,\qquad
e^{2A}=H^{-\fft{2}{\tilde d}}\,,\qquad e^{2B}=H^{\fft{2}{d}} \,,\cr
F_{(n)} &=& \fft{2}{\sqrt{\Delta}}{*d}H^{-1}\wedge dt\wedge
dx^1\wedge\cdots dx^{\tilde d-1}\,,\label{magneticp}
\end{eqnarray}
where $*$ denontes the Hodge dual and $H$ is a harmonic function in
the flat transverse space $ds_{D-\tilde d}^2 = dy^i dy^i$.  As in
the electric branes, the pseudo-Killing spinors can also be
obtained. They take the same form as (\ref{fakeksres1}) and
(\ref{fakeksres2}), except that now the projection operator is given
by
\begin{equation}
\Gamma \equiv \fft{{\rm i}^{[n/2]}}{(n-1)!} \epsilon_{\bar
i_1\cdots\bar i_{n-1}} \Gamma^{\bar i_1\cdots\bar i_{n-1}}\,.
\end{equation}

\section{Pseudo-supersymmetric intersecting branes}

For the right conditions, the $p$-brane solutions obtained in the
previous section can intersect with each other.  Here we shall
consider only the harmonic intersection where each ingredient
component is described by its harmonic function in a straightforward
way. Harmonic intersections of supersymmetric $p$-branes was
obtained in \cite{guev,tsey,pt}.  We shall present in the following
subsections the pair-wise intersections, from which all possible
multi-intersections can be built straightforwardly.

\subsection{Electric/electric intersection}

The full spacetime is split into four categories: $k$-dimensional
overall world volume $x^\mu$, $2(d-k)$-dimensional relative spaces
$(u^\alpha, v^\alpha)$ and the remaining $(D-2d+k)$-dimensional
overall transverse space $y^i$.  The ansatz is given by
\begin{eqnarray}
ds^2&=& (H_1H_2)^{-\fft{2}{d}} dx^\mu dx_\mu + H_1^{-\fft{2}{d}}
H_2^{\fft{2}{\tilde d}} du^\alpha du^\alpha + H_1^{\fft{2}{\tilde
d}} H_2^{-\fft{2}{d}} dv^\alpha dv^\alpha + (H_1H_2)^{\fft{2}{\tilde
d}} dy^i dy^i\,,\cr 
F_{(n)} &=& \fft{2}{\sqrt{\Delta}}\Big(dH_1^{-1}\wedge d^{(k)}
x\wedge d^{(d-k)} u + dH_2^{-1}\wedge d^{(k)} x\wedge d^{(d-k)} v
\Big)\,,
\end{eqnarray}
where $H_1$ and $H_2$ are harmonic functions of the $y$ space,
namely
\begin{equation}
\Box_{\vec y} H_{1}=0\,,\qquad \Box_{\vec y} H_2=0\,.
\end{equation}
This ansatz satisfies the full set of equations of motion of the
Lagrangian (\ref{genlag}) provided that
\begin{equation}
\fft{d^2}{D-2}=k=\hbox{positive integers}\,.\label{elcelccon}
\end{equation}
Applying this to M-theory or type IIB supergravity, we conclude that
the M2/M2 intersection intersection gives rise to a black hole
($k=1$) and D3/D3 intersection gives rise to a string $(k=1)$.  Although
there is an infinite number of solutions to (\ref{elcelccon}), the
dimensions $D$ rise rapidly. The next example is $(D,d,k)=(14,6,3)$,
corresponding to the 5/5-intersection in 16 dimensions that gives
rise to a 2-brane.

The pseudo-Killing spinors can also be calculated easily, which
satisfy the following two projections
\begin{equation}
\widetilde \Gamma \epsilon=\epsilon\,,\qquad \widehat \Gamma
\epsilon=\epsilon\,,
\end{equation}
where
\begin{equation}
\widetilde \Gamma=\Gamma^{01\cdots(k-1)}\Gamma^{k\cdots
(d-k-1)}\,,\qquad \widehat
\Gamma=\Gamma^{01\cdots(k-1)}\Gamma^{(d-k)\cdots (2d-k-1)}\,.
\end{equation}
For $\widetilde \Gamma$ and $\widehat \Gamma$ to have common
eigenvalues, they must commute, which implies that
\begin{equation}
d-k=\hbox{even}\,.
\end{equation}

\subsection{Electric/magnetic intersection}

The full spacetime is split into four categories: $\tilde
k$-dimensional overall world volume $x^\mu$, $(d+\tilde d-2\tilde
k)$-dimensional relative spaces $(u^\alpha, v^\beta)$ and the
remaining $(\tilde k+2)$-dimensional overall transverse space $y^i$.
The indices $\alpha$ and $\beta$ run $(d-\tilde k)$ and $(\tilde
d-\tilde k)$ values. The ansatz is given by
\begin{eqnarray}
ds^2&=& H_1^{-\fft2{d}}H_2^{-\fft{2}{\tilde d}} dx^\mu dx_\mu +
H_1^{-\fft{2}{d}} H_2^{\fft{2}{d}} du^\alpha du^\alpha +
H_1^{\fft{2}{\tilde d}} H_2^{-\fft{2}{\tilde d}} dv^\beta dv^\beta +
H_1^{\fft2{\tilde d}}H_2^{\fft{2}{d}} dy^i dy^i\,,\cr 
F_{(n)} &=& \fft{2}{\sqrt{\Delta}}\Big(dH_1^{-1}\wedge d^{(\tilde
k)} x\wedge d^{(d-\tilde k)} u + {*d}H_2^{-1}\wedge d^{(\tilde k)}
x\wedge d^{(\tilde d-\tilde k)} v\Big)\,,
\end{eqnarray}
where $H_1$ and $H_2$ are harmonic functions of the $y$ space. The
ansatz satisfies the full set of equations of motion of the
Lagrangian (\ref{genlag}) provided that
\begin{equation}
\fft{d\tilde d}{D-2}=\tilde k=\hbox{positive
integers}\,.\label{elcmagcon}
\end{equation}
Applying this to M-theory, we conclude that the M2/M5 intersection
intersection gives rise to a string ($\tilde k=2$).  There are two
independent projection gamma matrix operators associated with the
electric and magnetic brane components for pseudo-Killing spinors.
These two projectors have to commute, yielding to the following
additional condition
\begin{equation}
\tilde k={\rm even}\,.
\end{equation}

\subsection{Magnetic/magnetic intersection}

The full spacetime is split again into four categories: $\hat
k$-dimensional overall world volume $x^\mu$, $2(\tilde d-\hat
k)$-dimensional relative spaces $(u^\alpha, v^\alpha)$ and the
remaining $(D-2\tilde d+\hat k)$-dimensional overall transverse
space $y^i$. The ansatz is given by
\begin{eqnarray}
ds^2&=& (H_1H_2)^{-\fft{2}{\tilde d}} dx^\mu dx_\mu +
H_1^{-\fft{2}{\tilde d}} H_2^{\fft{2}{d}} du^\alpha du^\alpha +
H_1^{\fft{2}{d}} H_2^{-\fft{2}{\tilde d}} dv^\alpha dv^\alpha +
(H_1H_2)^{\fft{2}{d}} dy^i dy^i\,,\cr 
F_{(n)} &=& \fft{2}{\sqrt{\Delta}}{*\Big(dH_1^{-1}\wedge d^{(\hat
k)} x\wedge d^{(\tilde d-\hat k)} u + dH_2^{-1}\wedge d^{(\tilde k)}
x\wedge d^{(\tilde d-\hat k)} v \Big)}\,,
\end{eqnarray}
where $H_1$ and $H_2$ are harmonic functions of the $y$ space. The
ansatz satisfies the full set of equations of motion of the
Lagrangian (\ref{genlag}) provided that
\begin{equation}
\fft{\tilde d^2}{D-2}=\hat k=\hbox{positive
integers}\,.\label{magmagcon}
\end{equation}
Applying this to M-theory, we conclude that
the M5/M5 intersection gives rise to a 3-brane ($k=4$). As in the
previous examples, the existence of pseudo-Killing spinors requires an
additional condition
\begin{equation}
\tilde d-\hat k=\hbox{even}\,.
\end{equation}

It should be remarked that since $k + \tilde k=d$ and $\hat k+\tilde
k=\tilde d$.  The conditions for the possibility of harmonic
intersection for the electric/electric, electric/magnetic and
magnetic/magnetic cases are either satisfied for all or none.  The
same is the true for the existence of the pseudo-Killing spinors for
these solutions.  For the intersections with pseudo-supersymmetry,
the preserved fraction of the pseudo-Killing spinors is $\ft14$.

Having obtain the pair-wise intersections of the electric and
magnetic $p$-branes, it is straightforward to construct all possible
multi-intersections.

\section{Integrability conditions}

In the previous sections, we introduced pseudo-Killings spinors for
the theory (\ref{genlag}) and then obtained large classes of
solutions including $p$-branes and intersecting $p$-branes that
preserve a certain fraction of pseudo-Killing spinors.  In these
solutions, the equations of motion are solved without making use of
the pseudo-Killing spinor equations, but we verify that pseudo-Killing
spinors exist in these backgrounds.

The purpose of introducing pseudo-Killing spinors is to help us to
obtain new solutions that may otherwise not be possible to
construct. The pseudo-Killing spinor equations can reduce the
second-order equations to those of the first order.  However, the
success of obtaining pseudo-supersymmetric $p$-branes in earlier
sections should not give a wrong impression that the existence of a
pseudo-Killing spinor of a bosonic ansatz is sufficient for it to
satisfy the full set of equations of motion.  It is advantageous to
obtain the integrability condition for the pseudo-Killing spinor
equation.  The pseudo-Killing spinor equation together with the
integrability condition enables us to construct complicated new
solutions without having to verify explicitly the complicated
Einstein equations of motion.  We start with discussion for general
$(D,n)$ in the following subsection and then go on to specific cases
in the subsequent subsection.

\subsection{General analysis}

Here, we change the notation and express the pseudo-Killing spinor
equation as
\begin{equation}
D_M  \eta + b\left(\Gamma_M{}^{M_1M_2\dots
M_n}-a\delta_M^{M_1}\Gamma^{M_2\dots M_n}\right) F_{M_1M_2\dots
M_n}\eta=0\,.\label{ksnform}
\end{equation}
The constants $(a,b)$ can be read off from (\ref{ksdef1}) and
(\ref{ksdef11}), but are left arbitrary for now. The integrability
condition is given by
\begin{equation}
D_{[N}D_{M]}  \eta = T^\1_{NM} + T^\2_{NM}\,,\label{nformint1}
\end{equation}
where $T^\1$ and $T^\2$ are anti-symmetric tensors proportional to
the $n$-form field strength in linear and quadratic fashions
respectively,
\begin{eqnarray}
T^\1_{NM}&=& -b\left(\Gamma_{[M}\Gamma^{M_1M_2\dots
M_n}-(a+n)\delta_{[M}^{M_1}\Gamma^{M_2\dots M_n}\right)
\left(\nabla_{N]}F_{M_1M_2\dots M_n}\right)\eta\,,\cr T^\2_{NM}&=&
{b^2}\Big[(-1)^n\Gamma_{MN}\Gamma^{M_1M_2\dots
M_n}\Gamma^{N_1N_2\dots N_n}\cr
&&\qquad-(a+3n)(-1)^n\Gamma_{[M}\delta_{N]}^{M_1}\Gamma^{M_2\dots
M_n}\Gamma^{N_1N_2\dots N_n}\cr &&\qquad
-(a+n)\Gamma_{[M}\delta_{N]}^{N_1}\Gamma^{M_1M_2\dots M_n}
\Gamma^{N_2\dots N_n}\cr
&&\qquad-2(n-1)(a+n)(-1)^{n}\delta_{M}^{M_1}\delta_{N}^{M_2}\Gamma^{M_3\dots
M_n}\Gamma^{N_1N_2\dots N_n}\cr
&&\qquad+(a+n)^2\delta_{[M}^{M_1}\delta_{N]}^{N_1}\Gamma^{M_2\dots
M_n}\Gamma^{N_2\dots N_n} \Big]F_{M_1M_2\dots M_n}F_{N_1N_2\dots
N_n}\eta \,,
\end{eqnarray}
Contracting (\ref{nformint1}) with $\Gamma^N$, the left-hand side gives
\begin{equation}
4\Gamma^N D_{[N}D_{M]}\eta =R_{MN}\Gamma^{N}\,.\label{int1}
\end{equation}
The right-hand side gives much more complicated expressions
\begin{eqnarray}
\Gamma^N T^\1_{NM}&=&\ft{b}{2(n+1)} dF_{NM_1M_2\dots
M_n}\left(\Gamma_{M}{}^{NM_1M_2\dots M_n}
-\ft{n+1}na\delta_M^{[N}\Gamma^{M_1M_2\dots M_n]}\right)\eta
\cr&&+(-1)^{nD}\ft{b}2(*d*F)_{M_2\dots
M_n}\left(n\Gamma_{M}{}^{M_2\dots
M_n}-(n-1)a\delta_M^{M_2}\Gamma^{M_3\dots M_n}\right)\eta \cr&&
-\ft{b}2\left(\ft{n-1}na-D+n+1\right)\nabla_{M}F_{M_1M_2\dots
M_n}\Gamma^{M_1M_2\dots M_n}\eta\,,\label{int2}
\end{eqnarray}
and
\begin{eqnarray}
\Gamma^N T^\2_{NM}
\!\!\!\!&=&\!\!\!\!b^2\Big[\sum_{k=0}^{[\fft{n}{2}]}
\fft{(-1)^{\fft12(n+1)n-k}((n-1)!)^2}{(n-2k)!((2k)!)^2}
n[2ka-n(D-2k-n-1)]
\cr&&\qquad\times\delta^{M_1}_{N_1}\dots\delta^{M_{n-2k}}_{N_{n-2k}}
\Gamma_M{}^{M_{n-2k+1}\dots M_n}{}_{N_{n-2k+1}\dots N_n}F_{M_1\dots
M_n}F^{N_1\dots N_n} \cr&&+\sum_{k=0}^{[\fft{n-1}2]}
\fft{(-1)^{\fft12(n-2)(n-1)-k}((n-1)!)^2}{(n-2k-1)!(2k)!(2k+1)!} \cr
&&\times\Big((2k-n+1)a^2+4kna+n^2(n+2k-D+1)\Big)
\cr&&\times\delta^{M_2}_{N_2}\dots
\delta^{M_{n-2k}}_{N_{n-2k}}\Gamma^{M_{n-2k+1}\dots
M_n}{}_{N_1N_{n-2k+1}\dots N_n}F_{MM_2\dots M_n}F^{N_1N_2\dots N_n}
\cr&& +\sum_{k=0}^{[\fft{n}2]-1}
\fft{(-1)^{\fft12(n+1)n-k}((n-1)!)^2}{(n-2k-2)!(2k+1)!(2k+2)!} \cr
&& \times\Big((2k+2)a^2+n(2n-D)a-2n^2(k+1)\Big)
\cr&&\times\delta^{M_2}_{N_2}\dots
\delta^{M_{n-2k-1}}_{N_{n-2k-1}}\Gamma^{M_{n-2k}\dots
M_n}{}_{N_1N_{n-2k}\dots N_n}F_{MM_2\dots M_n}F^{N_1N_2\dots
N_n}\Big]\eta\,.\label{int3}
\end{eqnarray}
The vanishing of the last term in (\ref{int2}) requires that
\begin{eqnarray}
a=\fft{n(D-n-1)}{n-1}\,.\label{avalue}
\end{eqnarray}
Note that this condition fixes the relative sign and ratio between
$(\tilde\alpha,\tilde\beta)$ in (\ref{ksdef1}) and (\ref{ksdef11}).
Assemble the three structures (\ref{int1}), (\ref{int2}) and
(\ref{int3}), the term with single gamma matrix is
\begin{equation}
\Big[R_{MN}+4b^2n!(-1)^{\fft12(n+1)n}\Big(
(D-n-1)g_{MN}F^2+\big(\fft{1-n}{n}a^2+n(n-D+1)\big)F_{MN}^2\Big)\Big]
\Gamma^N\eta\,.
\end{equation}
The factor in the square-bracket is related to the Einstein's
equation of motion provided that
\begin{eqnarray}
b^2=(-1)^{\fft12(n+1)n}\fft{n-1}{8(D-2)(D-n-1)(n!)^2}\,.\label{bvalue}
\end{eqnarray}
Note that the pseudo-Killing spinor equation (\ref{ksnform}) with
the constants $(a,b)$ determined by (\ref{avalue}) and
(\ref{bvalue}) above is precisely the same one given in
(\ref{ksdef1}) and (\ref{ksdef11}), determined by examining the
vacuum and $p$-brane solutions. The integrability condition now
reduces to
\begin{eqnarray}
&&\Big[R_{MN}+\ft{n-1}{2(D-2)(n!)}g_{MN}F^2-\ft1{2(n-1)!}
F_{MN}^2\Big]\Gamma^N\eta\cr &&-\ft{2b}{n+1} dF_{NM_1M_2\dots
M_n}\left(\Gamma_{M}{}^{NM_1M_2\dots M_n}
-\ft{(n+1)(D-n-1)}{n-1}\delta_M^{[N}\Gamma^{M_1M_2\dots M_n]}\right)
\eta\cr &&-(-1)^{nD}2nb(*d*F)_{M_2\dots
M_n}\left(\Gamma_{M}{}^{M_2\dots
M_n}-(D-n-1)\delta_M^{M_2}\Gamma^{M_3\dots M_n}\right)\eta\cr
&&-\ft{1}{2(D-2)(D-n-1)}\sum_{k=1}^{[\fft{n}{2}]}
\ft{(-1)^{k}}{(n-2k)!((2k)!)^2}\Big[(2k-n+1)D+(n^2-4k-1)\Big]\cr
&&\qquad\times\delta^{M_1}_{N_1}\dots \delta^{M_{n-2k}}_{N_{n-2k}}
\Gamma_M{}^{M_{n-2k+1}\dots M_n}{}_{N_{n-2k+1}\dots N_n}F_{M_1\dots
M_n}F^{N_1\dots N_n}\eta\cr
&&-\ft{1}{2(D-n-1)(n-1)}\sum_{k=1}^{[\fft{n-1}2]}
\ft{(-1)^{k+1}}{(n-2k-1)!(2k)!(2k+1)!}
\Big[(2k-n+1)D+(n^2-4k-1)\Big]\cr
&&\qquad\times\delta^{M_2}_{N_2}\dots\delta^{M_{n-2k}}_{N_{n-2k}}
\Gamma^{M_{n-2k+1}\dots M_n}{}_{N_1N_{n-2k+1}\dots N_n}F_{MM_2\dots
M_n}F^{N_1N_2\dots N_n}\eta \cr\!\!\!&&\!\!\!
+\ft{D-2n}{2(D-2)(D-n-1)(n-1)}\sum_{k=0}^{[\fft{n}2]-1}
\ft{(-1)^{k}}{(n-2k-2)!(2k+1)!(2k+2)!}
\Big[(2k-n+3)D+(n^2-4k-5)\Big]\cr
&&\qquad\times\delta^{M_2}_{N_2}\dots
\delta^{M_{n-2k-1}}_{N_{n-2k-1}}\Gamma^{M_{n-2k}\dots
M_n}{}_{N_1N_{n-2k}\dots N_n}F_{MM_2\dots M_n}F^{N_1N_2\dots
N_n}\eta =0\,.
\end{eqnarray}
Thus we see that even if we substitute the full set of equations of
motion (\ref{formeom}-\ref{einsteineom}), there are still many terms
that are not vanishing.  These terms are quadratic in $F$, giving
rise to additional algebraic conditions for a bosonic configuration
to satisfy the equations of motion.  A careful analysis shows that
for the vacuum or $p$-brane solutions discussed earlier, these
algebraic constraints indeed vanish, as we would have expected.

Generically, in order for these extra contraints to vanish, we must
have, for all possible $k>0$, that
\begin{equation} D=\fft{n^2-4k-1}{n-2k-1}\,.
\end{equation}
This clearly cannot be satisfied in general. We may consider the
case where the above condition is satisfied for the lowest value of
$k$, namely $k=1$.  Then we have
\begin{equation}
D=\fft{n^2-5}{n-3}=n+3+\fft{4}{n-3}\,.
\end{equation}
The only integer solutions are $(n,D)=(4,11),(5,10),(7,11)$.  The
third case is Hodge dual to the first, and hence we shall not
consider it further.

It is worth mentioning that simplification occurs when $D=2n$ and
$n$ is odd. We can further impose the self-duality condition, the
integrability condition becomes
\begin{eqnarray}
&&\Big[R_{MN}-\fft1{2(n-1)!}F_{MN}^2\Big]\Gamma^N\eta
\cr&&-\fft{2b}{n+1} dF_{NM_1M_2\dots
M_n}\left(\Gamma_{M}{}^{NM_1M_2\dots M_n}
-(n+1)\delta_M^{[N}\Gamma^{M_1M_2\dots M_n]}\right)\eta
\cr&&-2nb(*d*F)_{M_2\dots M_n}\left(\Gamma_{M}{}^{M_2\dots
M_n}-(n-1)\delta_M^{M_2}\Gamma^{M_3\dots M_n}\right)\eta
\cr&&-\fft{1}{4(n-1)^2}\sum_{k=1}^{[\fft{n}{2}]}
\fft{(-1)^{k}}{(n-2k)!((2k)!)^2} \Big[2n(2k-n+1)+(n^2-4k-1)\Big]
\cr&&\qquad\times\delta^{M_2}_{N_2}\dots
\delta^{M_{n-2k}}_{N_{n-2k}}\Gamma^{M_{n-2k+1}\dots
M_n}{}_{N_1N_{n-2k+1}\dots N_n}F_{MM_2\dots M_n}F^{N_1N_2\dots
N_n}\eta=0\,.
\end{eqnarray}

\subsection{Critical dimensions for $n=5,4,3,2$}

In this section, we examine the integrability condition case by case
for $n=5,4,3,2$.

\bigskip
\noindent{$n=5$}
\medskip

The integrability condition for $n=5$ in arbitrary dimensions is given by
\begin{eqnarray}
&&\left[R_{MN}-\fft{1}{2(4!)}F^2_{MN}+\fft{2}{(5!)(D-2)}g_{MN}F^2\right]
\Gamma^N\eta \cr
&&-\ft{{\rm i}}{6!}\sqrt{\ft2{(D-6)(D-2)}}dF_{NM_1M_2M_3M_4M_5}
\Big(\Gamma_{M}{}^{NM_1M_2M_3M_4M_5}\!
-\!\ft53(D-6)\delta_M^{N}\Gamma^{M_1M_2M_3M_4M_5}\Big)\eta \cr
&&-\ft{{\rm i}}{4!}\sqrt{\ft2{(D-6)(D-2)}}(*d*F)_{M_1M_2M_3M_4}
\left(\Gamma_{M}{}^{M_1M_2M_3M_4}-(D-6)\delta_M^{M_1}
\Gamma^{M_2M_3M_4}\right)\eta\cr 
&& +\ft{2}{(5!)^2(D-6)(D-2)}
\Big[-100\Gamma_{M}{}^{M_2M_3M_4M_5N_2N_3N_4N_5}
F_{M_1M_2M_3M_4M_5}F^{M_1}{}_{N_2N_3N_4N_5}\cr
&&\qquad -300(D-10)\Gamma_{M}{}^{M_4M_5N_4N_5}F_{M_1M_2M_3M_4M_5}
F^{M_1M_2M_3}{}_{N_4N_5}\cr
&&\qquad+5(D-2)\Gamma^{M_2M_3M_4M_5N_1N_2N_3N_4N_5}
F_{MM_2M_3M_4M_5} F_{N_1N_2N_3N_4N_5}\cr
&&\qquad+100(D-10)\Gamma^{M_3M_4M_5N_1N_3N_4N_5}
F_{MM_2M_3M_4M_5}F_{N_1}{}^{M_2}{}_{N_3N_4N_5}\cr
&&\qquad+75(D-2)(D-10)\Gamma^{M_4M_5N_1N_4N_5}
F_{MM_2M_3M_4M_5}F_{N_1}{}^{M_2M_3}{}_{N_4N_5}\cr
&&\qquad+150(D-10)^2\Gamma^{M_5N_1N_5}F_{MM_2M_3M_4M_5}
F_{N_1}{}^{M_2M_3M_4}{}_{N_5}\Big]\eta=0\,.\label{n5int4}
\end{eqnarray}
Thus, we see that in general dimensions, the equations of motion are
necessary conditions for the integrability condition, but not
sufficient.  Additional constraints have to be imposed. As one would
have expected, these constraints are satisfied in the case of
pseudo-supersymmetric $p$-branes and intersecting branes discussed
in the previous sections.

    The integrability condition (\ref{n5int4}) clearly implies that
$D=10$ is the {\it critical} dimension for $n=5$, where many of the
constraints vanish.  For $D=10$, the integrability simplifies
significantly
\begin{eqnarray}
&&\left[R_{MN}-\fft{1}{2(4!)}F^2_{MN}+\fft{2}{(5!)(D-2)}
g_{MN}F^2\right]\Gamma^N\eta\cr
&&-\fft{{\rm i}}{4(6!)}dF_{NM_1M_2M_3M_4M_5}\left(
\Gamma_{M}{}^{NM_1M_2M_3M_4M_5}
-\fft{20}3\delta_M^{N}\Gamma^{M_1M_2M_3M_4M_5}\right)\eta
\cr&&-\fft{{\rm i}}{4(4!)}(*d*F)_{M_1M_2M_3M_4}\left(
\Gamma_{M}{}^{M_1M_2M_3M_4}-4\delta_M^{M_1}\Gamma^{M_2M_3M_4}
\right)\eta\cr&& 
+\fft{1}{4(4!)(5!)}\Big[-5\Gamma_{M}{}^{M_2M_3M_4M_5N_2N_3N_4N_5}
F_{M_1M_2M_3M_4M_5}F^{M_1}{}_{N_2N_3N_4N_5}\cr
&&\qquad+2\Gamma^{M_2M_3M_4M_5N_1N_2N_3N_4N_5}F_{MM_2M_3M_4M_5}
F_{N_1N_2N_3N_4N_5}\Big]\eta=0\,.
\end{eqnarray}

True Killing spinors arise when the 5-form is self-dual, for which
the theory becomes part of type IIB supergravity. In this case, we
have
\begin{eqnarray}
&&\epsilon^{MNM_2M_3M_4M_5N_2N_3N_4N_5}F_{M_1[M_2M_3M_4M_5}
F^{M_1}{}_{N_2N_3N_4N_5]}\cr 
&&\qquad \qquad \sim F^{[M}{}_{M_2M_3M_4M_5}F^{N]M_2M_3M_4M_5}=0\,,
\end{eqnarray}
as well as
\begin{eqnarray}
&&\Gamma^{M_2M_3M_4M_5N_1N_2N_3N_4N_5}F_{MM_2M_3M_4M_5}
F_{N_1N_2N_3N_4N_5}\eta\cr
&&\qquad\qquad=-5!F_{MM_2M_3M_4M_5}F_{N}{}^{M_2M_3M_4M_5}
\Gamma^{N}\eta\,.\label{9to1}
\end{eqnarray}
Here we have imposed the chirality condition
\begin{eqnarray}
\Gamma\eta=\eta\,,\qquad \Gamma = \Gamma^{012\cdots9}\,.
\end{eqnarray}
After imposing the chirality condition, the integrability condition becomes
\begin{eqnarray}
0&=&\left[R_{MN}-\fft{1}{4!}F^2_{MN}\right]\Gamma^N\eta
\cr&&-\fft{{\rm i}}{4(6!)}dF_{NM_1M_2M_3M_4M_5}
\left(\Gamma_{M}{}^{NM_1M_2M_3M_4M_5}
-\ft{20}3\delta_M^{N}\Gamma^{M_1M_2M_3M_4M_5}\right)\eta
\cr&&-\fft{{\rm i}}{4(4!)}({*d}{*F})_{M_1M_2M_3M_4}
\left(\Gamma_{M}{}^{M_1M_2M_3M_4}-4\delta_M^{M_1}
\Gamma^{M_2M_3M_4}\right)\eta\,.
\end{eqnarray}
This result was obtained in \cite{ggpr,gmsw}.  Thus for bosonic
ansatz with $dF=0=d{*F}$, the existence of a Killing spinor $\eta$
that gives rise to a time-like Killing vector, namely
\begin{equation}
\xi^M = \bar \eta \Gamma^M \eta\,,\qquad \xi^M\xi_M<0\,,
\end{equation}
implies the Einstein equations of motion.  It is of interest to
point out that the 9-gamma structure, which is dual to single-gamma
(\ref{9to1}), doubles the contribution of the energy-momentum tensor
associated with the self-dual 5-form.

    For our non-supersymmetric theory where the 5-form is not
self-dual, if the extra conditions
\begin{equation}
F_{M_1[M_2M_3M_4M_5}F^{M_1}{}_{N_2N_3N_4N_5]}=0\,,\qquad
F_{M[M_2M_3M_4M_5}F_{N_1N_2N_3N_4N_5]}=0\,.\label{n5extra}
\end{equation}
are satisfied, then the pseudo-Killing spinor equation plus
$dF=d*F=0$ will also automatically imply the Einstein equations of
motion. The conditions (\ref{n5extra}) can be easily satisfied if
the non-vanishing components of the 5-form are restricted to a
sub-manifold with dimensions less than or equal to seven.  This
enables us to construct new pseudo-symmetric bubbling AdS spaces in
section 7.

\bigskip
\noindent{$n=4$:}
\medskip

we find that the integrability condition becomes
\begin{eqnarray}
&&\left[R_{MN}-\fft1{12}\left(F^2_{MN}-\fft{3}{4(D-2)}
g_{MN}F^2\right)\right]\Gamma^N\eta\cr
&&-\fft1{2(5!)}\sqrt{\ft6{(D-5)(D-2)}}dF_{NM_1M_2M_3M_4}
\left(\Gamma_{M}{}^{NM_1M_2M_3M_4}-\ft53(D-5)\delta_M^{[N}
\Gamma^{M_1M_2M_3M_4]}\right)\eta\cr
&&-\ft1{2\sqrt{6(D-5)(D-2)}}(*d*F)_{M_1M_2M_3}
\left(\Gamma_{M}{}^{M_1M_2M_3}-(D-5)\delta_M^{M_2}
\Gamma^{M_3M_4}\right)\eta\cr
&&-\ft{4}{9}(D+7)b^2F_{[M_1M_2M_3M_4}F_{N_1N_2N_3N_4]}
\left(3\Gamma_{M}{}^{M_1M_2M_3M_4N_1N_2N_3N_4}\right.\cr
&&\qquad\qquad\left.
-4(D-8)\delta_{M}^{M_1}\Gamma^{M_2M_3M_4N_1N_2N_3N_4}
\right)\eta\cr 
&&+\ft{64}3b^2(D-11)(D-2)F_{MM_2M_3M_4}F_{N_1N_2N_3}
{}^{M_4}\Gamma^{M_2M_3N_1N_2N_3}\eta\label{n4genkscon}\\
&&+32b^2(D-11)F_{M_1M_2M_3M_4}F_{N_1N_2}{}^{M_3M_4}
\left(2(D-8)\delta_M^{M_1}\Gamma^{M_2N_1N_2}\!-\!
3\Gamma_{M}{}^{M_1M_2N_1N_2}\right)\eta=0\,.\nonumber
\end{eqnarray}
In general, the requirement of the vanishing of the last three terms
gives some additional algebraic constraints that are outside of the
equations of motion.

It is clear that the critical dimension for $n=4$ is $D=11$, for
which the last two terms vanish.  The third last term can now be
expressed as
\begin{equation}
\hbox{The third last term} =\ft{1}{72}*\left(F\wedge
F\right)_{M_5M_6M_7}\left[\Gamma_{M}{}^{M_5M_6M_7}-
6\delta_{M}^{M_5}\Gamma^{M_6M_7} \right]
\end{equation}
To derive the above identity at $D=11$, we have used
\begin{equation}
\Gamma^0\Gamma^1\cdots \Gamma^{10}=1\,,\qquad
\Gamma_{M_1\dots M_k}=\fft{(-1)^{\fft{k(k-1)}2}}{(11-k)!}
\epsilon_{N_1\dots N_{(11-k)}M_1\dots M_k}\Gamma^{N_1\dots N_{(11-k)}}\,.
\end{equation}
Thus the integrability condition for $D=11$ becomes \cite{fp,GP}
\begin{eqnarray}
0&=&\left(R_{MN}-\fft{1}{2(3!)}F^2_{MN}+\fft{1}{6(4!)}
g_{MN}F^2\right)\Gamma^N\eta\cr 
&&-\fft1{6!}dF_{NM_1M_2M_3M_4}\left(\Gamma_{M}{}^{NM_1M_2M_3M_4}
-10\delta_M^{[N}\Gamma^{M_1M_2M_3M_4]}\right)\eta\cr
&&-\ft{1}{36}*\left(d*F-\ft{1}{2}F\wedge F\right)_{M_1M_2M_3}
(\Gamma_{M}{}^{M_2M_3M_4}-6\delta_M^{M_2}
\Gamma^{M_3M_4})\eta\,.
\end{eqnarray}
This integrability condition implies that the proper equation of
motion for the 4-form is
\begin{equation}
d{*F}=\ft12F\wedge F\,.
\end{equation}
The origin of the self-interaction of the 4-form is the FFA term
\begin{equation}
{\cal L}_{FFA}=\ft16 F\wedge F\wedge A
\end{equation}
of eleven-dimensional supergravity. Thus we see that for $n=4$ in
$D=11$, the pseudo-Killing spinor is promoted to become the real
Killing spinor of $D=11$ supergravity.

For $D\ne 11$, the consistency between the existence of a
pseudo-Killing spinor and the equations of motion requires
additional constraints associated with the vanishing of the last
three terms in (\ref{n4genkscon}).  The AdS$\times$Sphere vacua, the
$p$-brane and intersecting $p$-branes discussed earlier satisfy
these constraints.  In section 8, we shall consider a specific
example in $D=8$ and demonstrate that the effect of these
constraints is to restrict severely the bubbling nature of the AdS
geometry.

\bigskip
\noindent{$n=3$:}
\medskip

We find that the integrability condition is given by
\begin{eqnarray}
&&\left[R_{MN}-\ft{1}{4}F^2_{MN}+\ft{1}{6(D-2)}g_{MN}F^2\right]
\Gamma^N\eta\cr 
&&-\ft1{4!}\sqrt{\ft1{(D-4)(D-2)}}dF_{NM_1M_2M_3}
\left(\Gamma_{M}{}^{NM_1M_2M_3}
-2(D-4)\delta_M^{[N}\Gamma^{M_1M_2M_3]}\right)\eta\cr
&&-\fft{(-1)^D}{2\sqrt{(D-4)(D-2)}}(*d*F)_{M_1M_2}
\left(\Gamma_{M}{}^{M_1M_2}-(D-4)\delta_M^{M_1}\Gamma^{M_2}\right)
\eta\cr
&&+72b^2F_{M_1[M_2M_3}F^{M_1}{}_{N_2N_3]}\Big[\Gamma_{M}{}^{M_2M_3N_2N_3}
-(D-6)\delta_{M}^{M_2}\Gamma^{M_3N_2N_3}\Big]\eta\cr 
&&-12b^2(D-2)\Gamma^{M_2M_3N_1N_2N_3}F_{MM_2M_3}F_{N_1N_2N_3}\eta=0
\,.\label{n3intcon}
\end{eqnarray}
In the above derivation, we have used
\begin{eqnarray}
F_{M_1[M_2M_3}F^{M_1}{}_{N_2N_3]}
&=&F_{M_1M_2[M_3}F^{M_1}{}_{N_2N_3]}\,.
\end{eqnarray}
The requirement of the vanishing of the three terms in the last two
lines in (\ref{n3intcon}) gives some additional algebraic
constraints.  The critical dimension is $D=6$, for which the
three-gamma term in the last two lines of (\ref{n3intcon}) vanishes.
In addition, we may impose chirality on the Killing spinor, namely
\begin{equation}
\Gamma\eta = -\eta\,,\qquad \Gamma=-\Gamma^0\Gamma^1\cdots\Gamma^5\,.
\end{equation}
We then have
\begin{eqnarray}
-12b^2(D-2)\Gamma^{M_2M_3N_1N_2N_3}F_{MM_2M_3}F_{N_1N_2N_3}\eta
=-\fft12F_{MM_2M_3}(*F)_{N}{}^{M_2M_3}\Gamma^{N}\eta\,.
\end{eqnarray}
Here we have used
\begin{equation}
\Gamma\Gamma_{M_1\dots
M_k}=-\fft{(-1)^{\fft{k(k-1)}2}}{(6-k)!}\epsilon_{N_1\dots
N_{(6-k)}M_1\dots M_k}\Gamma^{N_1\dots N_{(6-k)}}\,.
\end{equation}
If the 3-form is self dual, we have
\begin{equation}
\epsilon^{MNM_2M_3N_2N_3}F_{M_1[M_2M_3}F^{M_1}{}_{N_2N_3]}
=8F^{[M}{}_{M_2M_3}F^{N]M_2M_3}=0\,.
\end{equation}
Then the integrability condition becomes
\begin{eqnarray}
0&=&\left[R_{MN}-\ft12F^2_{MN}\right]\Gamma^N\eta
-\fft1{(4!)2\sqrt2}dF_{NM_1M_2M_3}\left(\Gamma_{M}{}^{NM_1M_2M_3}
-4\delta_M^{[N}\Gamma^{M_1M_2M_3]}\right)\eta
\cr&&-\fft1{4\sqrt{2}}(*d*F)_{M_1M_2}\left(\Gamma_{M}{}^{M_1M_2}
-2\delta_M^{M_1}\Gamma^{M_2}\right)\eta
\end{eqnarray}
This is precisely the integrability condition for ${\cal N}=(1,0)$
supergravity in six dimensions, studied in \cite{GMR}. The spinor
$\eta$ is promoted to be a real Killing spinor.  Bubbling
AdS$_3\times S^3$ solution were constructed in \cite{lv}, where an
additional axion has to be turned on in the $T^2$ direction.  Note
that the 5-gamma structure, which is dual to a single-gamma term,
doubles the contribution of the energy-momentum tensor associated
with the self-dual 3-form.

\bigskip
\noindent{$n=2$:}
\medskip

The integrability condition is
\begin{eqnarray}
&&\left[R_{MN}-\ft{1}{2}F^2_{MN}+\fft{1}{4(D-2)}g_{MN}F^2\right]
\Gamma^N\eta\cr
&&-\ft {\rm i}{3!\sqrt{2(D-3)(D-2)}}dF_{NM_1M_2}\left(\Gamma_{M}{}^{NM_1M_2}
-3(D-3)\delta_M^{[N}\Gamma^{M_1M_2]}\right)\eta \cr
&&-\ft{{\rm i}}{\sqrt{2(D-3)(D-2)}}(*d*F)_{M_1}\left(\Gamma_{M}{}^{M_1}
-(D-3)\delta_M^{M_1}\right)\eta\\
&&+ \ft{(D-1)}{48(D-3)(D-2)}(F\wedge F)_{M_1M_2N_1N_2}
\Big[\Gamma_{M}{}^{M_1M_2N_1N_2}-2(D-4)\delta_M^{M_1}
\Gamma^{M_2N_1N_2}\Big]\eta=0\,.\nonumber
\end{eqnarray}
Two critical dimensions arise in this case.  One is $D=4$, for which
the terms in the last line vanish.  The integrability condition is
then precisely that for ${\cal N}=2$, $D=4$ supergravity, namely
\begin{eqnarray}
&&\left[R_{MN}-\ft{1}{2}F^2_{MN}+\ft{1}{8}g_{MN}F^2\right]
\Gamma^N\eta-\ft{\rm i}{12}dF_{NM_1M_2}\left(\Gamma_{M}{}^{NM_1M_2}
-3\delta_M^{[N}\Gamma^{M_1M_2]}\right)\eta \cr
&&-\ft{{\rm i}}{2}(*d*F)_{M_1}\left(\Gamma_{M}{}^{M_1}
-\delta_M^{M_1}\right)\eta=0\,.
\end{eqnarray}

Another critical dimension is $D=5$, corresponding to $D=5$, ${\cal
N}=2$ supergravity, for which the integrability condition was
studied in \cite{gghpr}.  In $D=5$, we have
\begin{equation}
{\rm i}\Gamma^0\Gamma^1\cdots \Gamma^4=1\,,\qquad \Gamma_{M_1\dots
M_k}={\rm i}\fft{(-1)^{\fft{k(k-1)}2}}{(5-k)!}\epsilon_{N_1\dots
N_{(5-k)}M_1\dots M_k}\Gamma^{N_1\dots N_{(5-k)}}\,.
\end{equation}
The integrability condition can be further simplified, namely
\begin{eqnarray}
&&\left[R_{MN}-\ft12F^2_{MN}+\ft{1}{12}g_{MN}F^2\right]\Gamma^N\eta
\cr&&-\fft {\rm
i}{12\sqrt{3}}dF_{NM_1M_2}\left(\Gamma_{M}{}^{NM_1M_2}
-6\delta_M^{[N}\Gamma^{M_1M_2]}\right)\eta \cr&&-\fft {\rm
i}{2\sqrt{3}}\left[*(d*F+\fft1{\sqrt3}F\wedge
F)\right]_{M_1}\left(\Gamma_{M}{}^{M_1}-2\delta_M^{M_1}\right)\eta
=0\,.
\end{eqnarray}
This is very much like the case in $D=11$, and the equation of
motion for the 2-form is given by
\begin{equation}
d*F=-\ft1{\sqrt3} F\wedge F\,.
\end{equation}
The origin of this self-interaction of the 2-form is the FFA term
that needs to be augmented to the Lagrangian
\begin{equation}
{\cal L}_{FFA}=\ft1{3\sqrt3} F\wedge F\wedge A\,.
\end{equation}

\section{Pseudo-supersymmetric bubbling AdS in $D=10$}

As we have shown in the previous section, the integrability
conditions for the generic pseudo-Killing spinor equation imply that
additional constraints have to be imposed for a bosonic
configuration with pseudo-Killing spinors to satisfy the Einstein
equations of motion. For extremal $p$-branes, these extra conditions
are indeed satisfied.  In this section, we explore the possibility
of constructing bubbling AdS geometry based on the pseudo
supersymmetry.  The supersymmetric bubbling AdS geometries in type
IIB supergravity and M-theory were constructed in \cite{llm}. The
example we consider in this section is $n=5$ with $D=10$.  Of
course, if the 5-form is self-dual, the system is part of type IIB
supergravity, and was discussed in \cite{llm}.  We shall instead
consider an intrinsically non-supersymmetric theory where the 5-form
is not self-dual. The Lagrangian is given by (\ref{genlag}).  We are
looking for solutions with the $SO(4)\times SO(4)$ isometry, and the
most general ansatz is given by
\begin{eqnarray}
ds^2&=&g_{\mu\nu}dx^{\mu}dx^{\nu}+e^{H}d\Omega_3^2+e^{\tilde H}d\tilde\Omega_3^2\,,\cr
F_{_\5}&=&\ft12 F_{\mu\nu}dx^{\mu}\wedge dx^{\nu}\wedge \Omega_3
+\ft12{\tilde F}_{\mu\nu}dx^{\mu}\wedge dx^{\nu}\wedge \tilde\Omega_3
\,.\label{ads5bubble-ans}
\end{eqnarray}
The general pseudo-Killing spinor equation is given in section 2;
specializing to our specific case, we have
\begin{eqnarray}
0&=&D_M  \eta +\fft{{\rm i}}{960}\left(\Gamma_M{}^{M_1M_2M_3M_4M_5}-5\delta_M^{M_1}
\Gamma^{M_2M_3M_4M_5}\right)F_{M_1M_2M_3M_4M_5}\eta\cr
&=&D_M  \eta +\fft{{\rm i}}{960}\Gamma^{M_1M_2M_3M_4M_5}\Gamma_MF_{M_1M_2M_3M_4M_5}\eta\,.
\label{n5d10ks}
\end{eqnarray}
Note that this pseudo-Killing spinor equation takes the exact form
as the real one associated with the self-dual 5-form.

As was demonstrated in section 6, when the 5-form is not self-dual,
additional constraints (\ref{n5extra}) have to be imposed.  The
simplest way to satisfy the conditions (\ref{n5extra}) is to
restrict the non-vanishing components of the 5-form so that they all
lie only in 7 space-time directions. This can be achieved by setting
either $F_{\mu\nu}$ or $\tilde F_{\mu\nu}$ to zero.  Without loss of
generality we may set $\tilde F_{\mu\nu}=0$, and then the
non-vanishing components of $F_\5$ span on 4-dimensional space-time
with $g_{\mu\nu} dx^\mu dx^\nu$ and the $S^3$ with $d\Omega_3^2$,
totalling 7 directions.  This ansatz is different from the bubbling
AdS$_5$ in \cite{llm} where the self-duality of the 5-form is
required by type IIB supergravity and the $F$ and $\tilde F$ are
both non-vanishing.  Consequently, in the LLM solution, the
non-vanishing components of the self-dual 5-form lie in all 10
space-time directions.

We can now proceed and construct the new bubbling AdS$_5$ solution
supported by the non-self-dual 5-form.  The detail construction is
given in appendix \ref{appA}.  Here we shall just present the
solution:
\begin{eqnarray}
ds^2 &=& - h^{-2} (dt + V_i dx^i)^2 + h^2 (dy^2 + dx^idx^i) + y
e^{H} d\Omega_3^2 + y e^{-H} d \tilde \Omega_3^2 \cr h^{-2} &=& 2 y
\cosh H\,,\qquad y \partial_y V_i = \epsilon_{ij} \partial_j
D,\qquad y (\partial_i V_j-\partial_j V_i) = \epsilon_{ij}
\partial_y D\,,
  \cr
F_\5 &=& \Big[dB_t \wedge (dt + V) + h^{2}e^{3H} *_3d\Phi\Big]\wedge
d\Omega_3\,,\qquad
 B_t =\beta y^{2}e^{2 H}\,,\cr
 \Phi &=&-\alpha y^{2}e^{-2 H}\,,\qquad
 D =\ft12\alpha\beta \tanh H\,,\label{d10sol1}
\end{eqnarray}
where $\alpha,\beta=\pm1$, depending on the detail structure of the
pseudo-Killing spinor, and the basic function $D$ satisfies
\begin{eqnarray}
&&\partial^2_iD+y\partial_y\left(\fft1y\partial_yD\right)=0 \label{10dbeq}
\,.
\end{eqnarray}
To avoid the singularities at $y=0$, we must impose the following
boundary condition
\begin{eqnarray}
D(y=0)=\pm \ft12\,.\label{d10n5Dcon}
\end{eqnarray}
As in LLM case, the general solution of (\ref{10dbeq}) is given by
\begin{eqnarray}
D(x_1,x_2,y)&=&\frac{y^2}{\pi}\int_{\cal D} \frac{D(x_1',x'_2,0)
dx_1'dx_2'}{[({\bf x}-{\bf x}')^2+y^2]^2} =
-\frac{1}{2\pi}\int_{\partial {\cal D}} dl~
n_i'\frac{x_i-x_i'}{[({\bf x}-{\bf x}')^2+y^2]} +\sigma\,,\cr
V_i(x_1,x_2,y) &=& \fft{\epsilon_{ij}}{\pi} \int_{\cal D}
\frac{D(x_1',x'_2,0) (x_j - x'_j) dx_1'dx_2'}{[({\bf x}-{\bf
x}')^2+y^2]^2} =  \fft{\epsilon_{ij}}{2 \pi} \oint_{\partial {\cal
D}} \fft{dx'_j}{({\bf x}-{\bf x}')^2+y^2}\,.\label{integralsol}
\end{eqnarray}

The new solution we obtained shares the same characteristic
properties as the LLM solution.  The solution is completely fixed by
the boundary condition (\ref{d10n5Dcon}) on the $y=0$
two-dimensional plane.  The plus or minus choice indicates the
collapsing of either $S^3$ or $\tilde S^3$ respectively on the $y=0$
boundary. One important difference is that the 5-form field strength
given in (\ref{d10sol1}) can never be self-dual in our set up. In
particular, the 5-form carries either electric or magnetic fluxes
for the AdS$_5\times$S$^5$ geometry when the boundary condition is
in the shape of a round disc with either $S^3$ or $\tilde S^3$
shrinking inside respectively. Thus neither the theory nor the
solution can be embedded in type IIB supergravity.

For the LLM solution, the corresponding BPS states have simple field
theoretical description in terms of free fermions \cite{ber,cjr}.
The smooth geometric configurations are dual to the arbitrary
droplets of free fermions in the phase space \cite{llm}. In our
case, the dual field theory of the AdS boundary is an intrinsically
non-supersymmetric gauge theory; nevertheless, our construction
suggests that arbitrary free-fermion droplets can also exist in a
non-supersymmetric Yang-Mills theory.  The origin of these bubbling
states may lie in the $SO(6)$ global symmetry of the boundary
conformal field theory, which is the same as that of the
four-dimensional ${\cal N}=4$ super-conformal Yang-Mills theory.

\section{Pseudo-supersymmetric less-bubbling AdS in $D=8$}

The critical dimension for $n=4$ is $D=11$, as demonstrated in
section 6. For other dimensions, there can be no supersymmetry, and
the existence of the pseudo-Killing spinor of a bosonic ansatz does
not necessarily imply that Einstein's equations of motion are
satisfied.  Additional constraints have to be imposed.  In this
section, we investigate the effect of these constraints on bubbling
AdS solutions, by constructing an explicit solution for $D=8$.  As
can be seen from (\ref{n4genkscon}) that the constraints implies
that the following condition
\begin{equation}
F_{[MN}{}^{RS} F_{PQ]RS}=0\,.\label{d8constr}
\end{equation}

The detail construction can be found in appendix B. Here we shall
simply present the solution:
\begin{eqnarray}
ds^2 &=& - h^{-2}(dt + V_i dx^i)^2 + h^2 (dy^2 + dx^idx^i) + y e^{H}
d\Omega_2^2 + y e^{ - H} d \tilde \Omega_2^2\,,\cr F &=& (dB_t
\wedge (dt + V) + h^{2}e^{2H}{*_3d}\Phi)\wedge\Omega_\2\,,
\end{eqnarray}
where
\begin{eqnarray}
h^{-2} &=& 2 y \cosh H\,,\qquad
 y \partial_y V_i = \epsilon_{ij} \partial_j D,\qquad
 y (\partial_i V_j-\partial_j V_i) = \epsilon_{ij} \partial_y D
\,,\cr
B_t &=&-\fft{2m}{\sqrt 3}y^{\fft32}e^{\fft32 H}\,,\qquad
\Phi =-\fft{2}{\sqrt 3}y^{\fft32}e^{-\fft32 H}\,,\cr
D &=&-m \tanh H   \label{solmetric4}\,.
 \end{eqnarray}
Here $m=\pm1$ and the basic function $D$ satisfies
\begin{eqnarray}
&&\partial^2_iD+y\partial_y\left(\fft1y\partial_yD\right)=0 \,,\label{8Dbeq}\\
&&(\partial_{i}D)^2+(\partial_y D)^2=y^{-2}(1-D^2)^2\,.\label{8Dconstr}
\end{eqnarray}
To avoid the singularities at $y=0$, we must impose the following
boundary condition
\begin{eqnarray}
D(y=0)=\pm1\,.\label{d8boundcon}
\end{eqnarray}
The solution for (\ref{8Dbeq}) is given in (\ref{integralsol}). Note
that the additional equation (\ref{8Dconstr}) is a consequence of
imposing the condition (\ref{d8constr}).  Note that the constraint
(\ref{8Dconstr}) is invariant under $D\leftrightarrow 1/D$.

The non-linearity of (\ref{8Dconstr}) implies that we can no longer
generate new solutions by superposing the known solutions.  This
severely restricts the bubbling effect that is associated with
(\ref{8Dbeq}). We have tested boundary conditions for many shapes,
such as the disc, rectangle, ring, belt and multi-discs. The only
solutions we have found analytically are the AdS$_4\times S^4$ and
pp-wave solutions, which correspond to the boundary conditions in
the shapes of a disc and half-filled plane respectively.  On the
other hand, the constraint (\ref{8Dconstr}) is clearly consistent
with the boundary condition (\ref{d8boundcon}). It is quite possible
that less-bubbling solutions beyond the vacuum solution may exist.
These solutions would correspond to some specific free-fermion
droplets in the phase space of the dual gauge theory.

\section{Conclusions}

In this paper we consider Einstein gravity coupled to an $n$-form
field strength in general $D$ dimensions. We introduce the
pseudo-Killing spinor equation for such a system. In the special
case when the system becomes (part of) the bosonic Lagrangian of a
supergravity theory, the pseudo-Killing spinors become real Killing
spinors.  We show by explicit construction, for AdS$\times$Sphere
vacuum solution, extremal $p$-branes and intersecting $p$-branes,
pseudo-Killing spinors behave just like real Killing spinor in
supergravities.  The vacua have the maximum number of pseudo-Killing
spinors whilst $p$-branes and intersecting $p$-branes have fractions
of pseudo-Killing spinors.

We study the integrability condition of the pseudo-Killing spinor
equation. We find that additional constraints have to be imposed so
that the bosonic ansatz with pseudo-Killing spinors can
automatically satisfy the Einstein equations of motion.  For
$n=5,4,3,2$, there exist critical dimensions for which the
additional constraints vanish; these corresponds to relevant
supergravities.  However, in some non-supersymmetric cases,
additional constraints can nevertheless be satisfied by appropriate
ansatz, leading to the construction of new non-trivial
pseudo-supersymmetric solutions with pseudo-Killing spinors. Thus,
even though the pseudo-Killing spinors we introduced in this paper
are not consistent at the full level with those non-supersymmetric
theories, they provide a useful technique for reducing second-order
Einstein equations to the first-order system, for a large number of
special cases of ansatze that can circumvent the additional
constraints. Even in the case when the additional constraints cannot
be circumvented in a trivial way, the technique can still provide
non-trivial solutions. These solutions are unlikely to be found by
considering equations of motion only.

A concrete example we present is the new bubbling AdS geometry in
$D=10$ supported by a non-self-dual 5-form.  Although the solution
exhibit almost identical features of the previously-constructed LLM
solution, it is intrinsically non-supersymmetric and cannot be
embedded in type IIB supergravity in any limit of parameters of the
solution.  This demonstrates that bubbling states of arbitrary
free-fermion droplets in the phase space can exist also in
non-supersymmetric boundary conformal field theory.  The origin of
these states are not due to the supersymmetry, but likely to be
associated with the $SO(6)$ global symmetry, which is shared by both
the supersymmetric and non-supersymmetric conformal field theories.
The bubbling geometry we constructed preserves half of
pseudo-supersymmetry. It is of interest to investigate whether there
are bubbling geometry with less pseudo-supersymmetry, analogous to
the $\ft14$- and $\ft18$-BPS solutions constructed in
\cite{ccdlllvw}.

We also present a concrete example of $(D,n)=(8,4)$ for which the
additional constraint can not be circumvent.  This additional
constraint gives rise to non-linear differential equation on the
basic function of the would-be bubbling geometry.  The non-linearity
implies that we cannot construct new solutions by superposing the
known solutions.  This severely restricts the bubbling nature of the
geometry. Free-fermion droplets in the dual gauge theory can no
longer arbitrary. Nevertheless there may still exist less bubbling
configurations that go beyond the vacuum solution.  How to solve
this non-linear equation remains an open problem.

Thus our introduction of pseudo-Killing spinors to an intrinsically
non-supersymmetric system can help us to construct new solutions
that are unlikely to be found by examining only the equations of
motion.  This technique enlarges the possibility of constructing
more non-trivial bulk gravity backgrounds.  Applying the AdS/CFT
correspondence, our results show that an intrinsically
non-supersymmetric conformal field theory may have
pseudo-supersymmetric states with characteristics of BPS states of
superconformal field theory.

\section*{Acknowledgement}

H.L.~is grateful to the Mitchell Institute for Fundamental Physics
and Astronomy at Texas A\&M University for hospitality during the
course of this work.  We are grateful to Haishan Liu for reading the
manuscript and correcting the typos.

\appendix

\section{Detail derivation of the new bubbling AdS$_5$ solution}
\label{appA}

In this appendix, we provide detail derivation of the new bubbling
AdS$_5\times S^5$ geometry, presented in section 7.  The Lagrangian
is given by (\ref{genlag}), but specialized to $n=5$ and $D=10$.
This is an intrinsically non-supersymmetric theory since we do not
require that the 5-form be self dual. The pseudo-Killing spinor
equation is given by (\ref{n5d10ks}). Of course, they become real
Killing spinors if we impose the self-duality for the 5-form.  The
ansatz with the $SO(4)\times SO(4)$ isometry is given by
(\ref{ads5bubble-ans}). Note that we use $\mu$, $a$ and $\tilde a$
to denote indices for $ds_4^2$, $d\Omega_3^2$ and
$d\tilde\Omega_3^2$ respectively.

To proceed, we decompose the gamma matrices as follows
\begin{eqnarray}
\Gamma_{\mu}=\gamma_{\mu}\otimes 1\otimes 1\otimes
1\,,\qquad\Gamma_a= \hat\gamma\otimes \sigma_{\hat 1} \otimes
\sigma_a\otimes 1\,,\qquad\Gamma_{\tilde a}= \hat\gamma \otimes
\sigma_{\hat 2}\otimes 1\otimes \sigma_{\tilde a}
\end{eqnarray}
where
\begin{eqnarray}
\hat\gamma ={\rm i}\gamma_{\hat0\hat1\hat2\hat3}=-\fft1{4!}{\rm
i}\epsilon_{\mu_1\mu_2\mu_3\mu_4}
 \gamma^{\mu_1\mu_2\mu_3\mu_4}\,,\qquad\epsilon_{\hat0\hat1\hat2\hat3}=1
 \,,\qquad\sigma_{\hat 3}=-{\rm i}\sigma_{\hat1}\sigma_{\hat2}\,.
\end{eqnarray}
Here the hatted indices are the vielbein indices.  The $\sigma_{\hat
i}$'s are the Pauli matrices and the hatted indices in Gamma
matrices are the vielbein indices. The $\Gamma_{11}$ in this
decomposition is given by
\begin{equation}
\Gamma_{11}=\hat \gamma\,\sigma_{\hat 3}\,.
\end{equation}

We now perform the reduction on $S^3\times S^3$ by introducing two
component spinors $\chi_{\pm}$ and $\tilde\chi_{\pm}$ which obey the
equations
\begin{equation}\label{SpinorOnS3}
{\nabla}'_a \chi_\alpha=\alpha\frac{{\rm i}}{2}{\sigma}_{\hat
a}\chi_\alpha\,,\qquad{\nabla}'_{\tilde a} \tilde
\chi_\beta=\beta\frac{{\rm i}}{2}{\sigma}_{\hat{\tilde a}}\tilde
\chi_\beta\,.
\end{equation}
Note that $(\alpha,\beta)$ are $\pm 1$ when they appear in the
equations as numbers and $\pm$ as indices of the Killing spinors
$\chi_\pm$ and $\tilde \chi_\pm$. The ${\nabla}'$ is the covariant
derivative on a unit sphere. It is related to covariant derivatives
in the sphere directions in our ansatz as
\begin{equation} \label{spincon1}
\nabla_a={\nabla'}_a-\ft{1}{4}{\Gamma^\mu}_a\partial_\mu H,\qquad
\nabla_{\tilde a}={\nabla'}_{\tilde a}-
\ft{1}{4}{\Gamma^\mu}_{\tilde a}\partial_\mu \tilde H\,.
\end{equation}

We may expand the spinor $\eta$ over basis of the real Killing
spinors on the two spheres
\begin{equation}\label{DecompSpinXi}
\eta=\sum_{\alpha,\beta=+,-}\epsilon_{\alpha\beta}
\otimes\chi_{\alpha}\otimes\tilde \chi_{\beta}\,.
\end{equation}
All pseudo-Killing spinors can be written as a linear combination of
chiral and anti-chiral spinors, defined by
\begin{equation}
\Gamma_{11}\eta=\lambda\eta\,,\qquad \lambda=\pm 1
\end{equation}
Since the $\Gamma_{11}$ commutes with the Killing spinor equation,
the chiral and anti-chiral Killing spinors are independent of each
other. For simplicity, we do not use any subscript on $\eta$ to
distinguish the two types of pseudo-Killing spinors, but instead
just let the unspecified parameter $\lambda$ to indicate the
chirality of the spinor.  By contrast, in type IIB supergravity with
the self-dual 5-form, the chirality is specified. The expression in
(\ref{n5d10ks}) involving the 5-form becomes
\begin{eqnarray}
&&\fft{{\rm i}}{960}\Gamma^{M_1M_2M_3M_4M_5} F_{M_1M_2M_3M_4M_5}
\cr& =&\fft{{\rm i}}{96} (
e^{-\fft32H}F_{\mu\nu}\Gamma^{\mu\nu}\epsilon_{abc}\Gamma^{abc} +
e^{-\fft32 \tilde H }{\tilde F}_{\mu\nu}\Gamma^{\mu\nu}
\epsilon_{{\tilde a}{\tilde b}{\tilde c}}\Gamma^{{\tilde a}{\tilde
b}{\tilde c}}) \cr& =&-\fft1{16} (
e^{-\fft32H}F_{\mu\nu}\gamma^{\mu\nu}\hat\gamma\otimes \sigma_{\hat
1}\otimes1\otimes1 +  e^{-\fft32 \tilde H }{\tilde
F}_{\mu\nu}\gamma^{\mu\nu}\hat\gamma\otimes \sigma_{\hat
2}\otimes1\otimes1)\,.
\end{eqnarray}
The pseudo-Killing spinor equations in $(\mu, a, \tilde a)$ directions
are given by
\begin{eqnarray}
&&\sum_{\alpha,\beta}\Big[\nabla_{\rho}\epsilon_{\alpha\beta}
\otimes\chi_{\alpha}\otimes\tilde \chi_{\beta}-\ft1{16} \Big(
e^{-\fft32H}F_{\mu\nu}\gamma^{\mu\nu}\hat\gamma\gamma_{\rho}
\sigma_{\hat
1}\epsilon_{\alpha\beta}\otimes\chi_{\alpha}\otimes\tilde
\chi_{\beta}\cr &&\qquad\qquad\qquad +  e^{-\fft32 \tilde H }{\tilde
F}_{\mu\nu}\gamma^{\mu\nu}\hat\gamma\gamma_{\rho}\sigma_{\hat 2}
\epsilon_{\alpha\beta}\otimes\chi_{\alpha}\otimes\tilde
\chi_{\beta}\Big)\Big]=0 \,,\cr &&\sum_{\alpha,\beta}\left[\fft{{\rm
i}\alpha}2e^{-\fft{H}2} (\epsilon_{\alpha\beta}\otimes
\sigma_a\chi_{\alpha}\otimes\tilde
\chi_{\beta})-\ft14\partial_{\mu}H(\gamma^{\mu} \hat\gamma
\sigma_{\hat 1}\epsilon_{\alpha\beta}\otimes
\sigma_a\chi_{\alpha}\otimes\tilde \chi_{\beta}) \right.\cr&&\left.
-\fft1{16} (
e^{-\fft32H}F_{\mu\nu}\gamma^{\mu\nu}\epsilon_{\alpha\beta}\otimes
\sigma_a\chi_{\alpha}\otimes\tilde \chi_{\beta} +{\rm i}  e^{-\fft32
\tilde H }{\tilde F}_{\mu\nu}\gamma^{\mu\nu}\sigma_{\hat
3}\epsilon_{\alpha\beta} \otimes\sigma_a\chi_{\alpha}\otimes\tilde
\chi_{\beta})\right]=0\,, \cr &&\sum_{\alpha,\beta}\left[\fft{{\rm
i}\beta}2e^{-\fft{\tilde H}2}(\epsilon_{\alpha\beta}\otimes
\chi_{\alpha}\otimes\sigma_{\tilde a}\tilde \chi_{\beta})
-\ft14\partial_{\mu}{\tilde H}(\gamma^{\mu} \hat\gamma\hat
\sigma_2\epsilon_{\alpha\beta}\otimes \chi_{\alpha}\otimes
\sigma_{\tilde a}\tilde \chi_{\beta})\right.\label{kssum}\\
&&\left. -\fft1{16} (-{\rm i}
e^{-\fft32H}F_{\mu\nu}\gamma^{\mu\nu}\hat
\sigma_3\epsilon_{\alpha\beta}\otimes\chi_{\alpha}\otimes\sigma_{\tilde
a}\tilde \chi_{\beta} +  e^{-\fft32 \tilde H }{\tilde
F}_{\mu\nu}\gamma^{\mu\nu}\epsilon_{\alpha\beta}\otimes\chi_{\alpha}\otimes
\sigma_{\tilde a}\tilde \chi_{\beta})\right]=0\,.\nonumber
\end{eqnarray}
It is easy to see that the spinor basis $\chi_\alpha\otimes
\tilde\chi_\beta$ in the decomposition (\ref{DecompSpinXi}) are
independent. Thus there are four independent possible solutions
depending on the discrete choices of the $\pm 1$ values of the
parameters $\alpha$ and $\beta$.  However, they cannot be
simultaneous solutions. If we choose one set of $(\alpha,\beta)$
values to obtain pseudo-Killing spinors, the other three choices
cannot lead to any solutions.  Thus we can drop the sum notation in
the equations in (\ref{kssum}). Furthermore, we shall drop the
subscripts for $\epsilon_{\alpha\beta}$ for convenience from now on.
It should be understood that the parameters $(\alpha,\beta)$ now
denote one, but unspecified choice.  This leads to
\begin{eqnarray}
&&\left[\nabla_{\rho} -\ft1{16} (
e^{-\fft32H}F_{\mu\nu}\gamma^{\mu\nu}\hat\gamma\gamma_{\rho}\sigma_{\hat
1} +  e^{-\fft32 \tilde H }{\tilde
F}_{\mu\nu}\gamma^{\mu\nu}\hat\gamma\gamma_{\rho}\sigma_{\hat
2})\right]\epsilon=0 \,,\cr &&\left[{\rm i}\alpha
e^{-\fft{H}2}\hat\gamma\sigma_{\hat 1}
+\ft12\partial_{\mu}H\gamma^{\mu} -\ft1{8} (
e^{-\fft32H}F_{\mu\nu}\gamma^{\mu\nu}\hat\gamma\sigma_{\hat 1} -
e^{-\fft32 \tilde H }{\tilde
F}_{\mu\nu}\gamma^{\mu\nu}\hat\gamma\sigma_{\hat
2})\right]\epsilon=0\,, \cr &&\left[{\rm i}\beta e^{-\fft{\tilde
H}2}\hat\gamma\hat \sigma_2 +\ft12\partial_{\mu}{\tilde
H}\gamma^{\mu} +\ft1{8} (
e^{-\fft32H}F_{\mu\nu}\gamma^{\mu\nu}\hat\gamma\hat \sigma_1 -
e^{-\fft32 \tilde H }{\tilde
F}_{\mu\nu}\gamma^{\mu\nu}\hat\gamma\hat
\sigma_2)\right]\epsilon=0\,.
\end{eqnarray}
As was demonstrated in section 6, the existence of pseudo-Killing
spinors is not sufficient for the ansatz to satisfy the Einstein
equations of motion. Additional constraints (\ref{n5extra}) have to
be satisfied. For our ansatz, the constraints become
\begin{equation}
F_{\mu_1[\mu_2}\tilde F^{\mu_1}{}_{\nu_2]}=0\,,\qquad
F_{\mu_1[\mu_2}\tilde F_{\nu_1\nu_2]}-\tilde F_{\mu_1[\mu_2}
F_{\nu_1\nu_2]}=0\,,\qquad F_{[\mu_1\mu_2}\tilde
F_{\nu_1\nu_2]}=0\,.
\end{equation}
These constraints are difficult to solve except for two special
cases: one is that $F_\5$ is self-dual, and the other is that one of
the $F_\2$ and $\tilde F_\2$ vanishes. The former case was discussed
in \cite{llm}. We shall focus on the later case. Let $F_{(2)}=dB$
and $\tilde F_{(2)}=0$, the pseudo-Killing spinor equations now
become significantly simpler:
\begin{eqnarray}
&&\left[\nabla_{\rho}-\ft1{16}
e^{-\fft32H}F_{\mu\nu}\gamma^{\mu\nu}\hat\gamma\gamma_{\rho}\sigma_{\hat
1}\right]\epsilon=0
\,,\label{10dbeq1}\\
&&\left[{\rm i}\alpha e^{-\fft{H}2}\hat\gamma\sigma_{\hat 1}
+\ft12\partial_{\mu}H\gamma^{\mu}
-\ft1{8} e^{-\fft32H}F_{\mu\nu}\gamma^{\mu\nu}\hat\gamma
\sigma_{\hat 1}\right]\epsilon=0\,,\label{10dbeq2}
\\
&&\left[{\rm i}\beta e^{-\fft{\tilde H}2}\hat\gamma\hat \sigma_2
+\ft12\partial_{\mu}{\tilde H}\gamma^{\mu} +\ft1{8}
e^{-\fft32H}F_{\mu\nu}\gamma^{\mu\nu}\hat\gamma\hat
\sigma_1\right]\epsilon=0\,.\label{10dbeq3}\,.
\end{eqnarray}

We are now in the position to derive the solution using the standard
G-structure technique. Let us define the following real spinor
bilinears
\begin{eqnarray}
&& f_1={\rm i}{\bar\epsilon}\sigma_{\hat 1}\epsilon\,,\quad
 f_2={\rm i}{\bar\epsilon} \sigma_{\hat 2} \epsilon\,,\quad
K_\mu={\bar\epsilon}\gamma_\mu\epsilon\,,\quad
L_\mu={\bar\epsilon}\gamma_\mu \hat\gamma \epsilon\,, \quad
Y_{\mu\nu}={\bar\epsilon}\gamma_{\mu\nu}\sigma_{\hat 1}\epsilon\,,
\end{eqnarray}
as well as the complex spinor bilinears
\begin{equation}
L^c_\mu={\bar\epsilon}^c\gamma_\mu\sigma_{\hat 1}\epsilon\,,
\qquad Y^c_{\mu\nu}={\bar\epsilon}^c\gamma_{\mu\nu}\epsilon\,.
\end{equation}
where $\bar\epsilon=\epsilon^+\gamma^{\hat 0}$ and
$\bar\epsilon^c=\epsilon^t C$. We choose the basis where
$\gamma_{\hat 2}$ and $\sigma_{\hat 2}$ are antisymmetric while
other  $\gamma_{\hat\mu}$'s and $\sigma_{\hat 1}$ are symmetric,
thus $C=\gamma_{\hat 2} \sigma_{\hat 1}$. From (\ref{10dbeq}) we
have the following relations between the bi-spinors
\begin{eqnarray}
\nabla_{\mu}f_1&=& =\ft{1}{8} e^{- \fft32 H
}\epsilon_{\mu}{}^{\nu\rho\sigma}  F_{\nu\rho}K_{\sigma}\,,
\label{10ddf1}
\\
\nabla_{\mu}f_2&=&
\ft14\lambda e^{- \fft32 H }F_{\mu\nu} K^{\nu}\,,\label{10ddf2}
\\
\nabla_{\mu}K_{\nu}&=&-\ft{1}{8} e^{- \fft32 H }
\left(\epsilon_{\mu\nu}{}^{\lambda\rho}F_{\lambda\rho}f_1
+2\lambda F_{\mu\nu} f_2\right)\,,\label{10ddk}
\\
\nabla_{\mu}L_{\nu}&=&\ft{1}{8} e^{-\fft32  H }
\left(g_{\mu\nu}F_{\lambda\rho}Y^{\lambda\rho}
+2F_{\mu\rho}Y^{\rho}{}_{\nu}+2F_{\nu\rho}Y^{\rho}{}_{\mu}\right)\,,
\\
\nabla_{\mu}L^c_{\nu}&=&-\ft{1}{8} e^{-  H }
\left(g_{\mu\nu}F^{\lambda\rho}Y^c_{\lambda\rho}
+2F_{\mu}{}^{\rho}Y^{c}_{\rho\nu}+2F_{\nu}{}^{\rho}Y^{c}_{\rho\mu}\right)\,.
\end{eqnarray}
From (\ref{10ddk}), we have $\nabla_{(\mu} K_{\nu)}=0$.  Thus
$K=K^{\mu}\partial_\mu$ is a Killing vector.  Also the 1-forms
$L=L_{\mu}dx^{\mu}$ and $L^c=L^c_{\mu}dx^{\mu}$ are (locally) exact.
Using Fierz identities we obtain the following two identities
\begin{equation}
K\cdot L=0\,,\qquad
L^2=-K^2=f_1^2+f_2^2\,.
\end{equation}
Thus the Killing vector $K$ is time-like.

Since $L=L_{\mu}dx^{\mu}$ is (locally) exact, we can choose a
coordinate $y$ such that
\begin{equation}
  dy=L_\mu dx^\mu\,.
\end{equation}
We then choose the other three coordinates in the subspace orthogonal to $y$
\begin{equation}
ds^2=h^2 dy^2+g_{\alpha\beta}dx^\alpha dx^\beta\,,\qquad
h^{-2}=L^2=f_1^2+f_2^2\,.
\end{equation}
Using the relation
\begin{equation}
0=K^\mu L_\mu =  K^y L_y=  K^y\,,
\end{equation}
we find that $K^\alpha$ is a vector in three dimensional space
spanned by $x^\alpha$.  Since $K$ is time-like, we can choose the
time $t$ as the coordinate along $K^\alpha$, {\it i.e.}
$K=\partial/\partial_t$. We find
\begin{equation}
ds^2=-h^{-2}(dt+V_i dx^i)^2+h^2( dy^2+\tilde g_{ij}dx^i dx^j)
\end{equation}
where $i,j=1,2$.

The equation of motion for the 5-form is $d *_{10}F_{\5}=0$. For our
ansatz, this becomes
\begin{equation}\label{10dFldStrEqn}
d(e^{\fft32(\tilde H-H)}*_4 dB)=0\,.
\end{equation}
Let us write the components of the gauge field $B$ as
\begin{equation}
B_\mu dx^\mu=B_t dt+B_\alpha dx^\alpha=B_t(dt+V_idx^i)+(B_\alpha-B_t
V_\alpha) dx^\alpha\equiv B_t(dt+V_idx^i)+{\hat B}\,,
\end{equation}
then the components of the field strength and its dual are given by
\begin{eqnarray}
F&=&dB_t\wedge(dt+V)+(d{\hat B}+B_t dV)\,,\cr
*_4 F&=& h^{2}*_3 dB_t+h^{-2} (dt+V)\wedge*_3(d{\hat B}+B_t dV)\,,
\end{eqnarray}
where $*_3$ is the Hodge dual in $ds_3^2=dy^2+\tilde g_{ij}dx^i
dx^j$. The equation (\ref{10dFldStrEqn}) now becomes
\begin{equation}
d*_3\left[h^{-2}e^{\fft32(\tilde H-H)}(d{\hat B}+B_t dV)\right]=0\,.
\end{equation}
Thus, locally we can introduce a dual potential $\Phi$:
\begin{equation}\label{10dDefPhiM}
d{\hat B}+B_t dV=h^2\,e^{\fft32(H-\tilde H)}*_3d\Phi\,.
\end{equation}
The time component of (\ref{10dFldStrEqn}) leads to the equation:
\begin{equation}
d\left[V\wedge d\Phi+ h^2e^{\fft32(\tilde H-H)}*_3 dB_t\right]=0\,.
\end{equation}
From the equation (\ref{10ddf2}) and the fact that
$B_i$ is independent of $t$, we find
\begin{equation}
\partial_\mu f_2=\ft{1}{4}\lambda e^{-\fft32H}\partial_\mu B_t\,.
\end{equation}
By using (\ref{10dbeq2}),  we find
\begin{eqnarray}
\partial_\mu B_t&=& F_{\mu\nu} K^\nu=
 \ft{1}{4} \bar \epsilon [\gamma_\mu,\, {\not F}] \epsilon
\cr&=&\bar \epsilon [\gamma_\mu,\, 2i\alpha
e^{H}-e^{\fft32H}\partial_{\nu}H\gamma^{\nu} \hat\gamma\sigma_{\hat
1} ] \epsilon =2\lambda e^{\fft32H}\partial_{\mu}H f_2
\end{eqnarray}
Thus
\begin{equation}
\partial_\mu f_2=\ft12 f_2\partial_{\mu}H\quad \Rightarrow\quad
f_2= c_2 e^{\fft12H}\,,\qquad B_t= \lambda c_2e^{2 H}\,.
\end{equation}
From the equation (\ref{10ddf1}), we find
\begin{equation}
\partial_\mu f_1=\ft{1}8 e^{- \fft32 H }\epsilon_{\mu}{}^{\nu\rho\sigma}
F_{\nu\rho}K_{\sigma} =\ft{1}4 e^{- \fft32 H }(*_4F)_{\mu\nu}K^{\nu}
=-\ft{1}{4}e^{-\fft32\tilde H} \partial_{\mu}\Phi\,.
\end{equation}
By using (\ref{10dbeq3}), we have
\begin{eqnarray}
\epsilon_{\mu\nu\rho\sigma}F^{\nu\rho}K^\sigma&=& \fft{{\rm i}}{2}
F_{\nu\rho} {\bar\epsilon} \hat\gamma
\{\gamma_{\mu},\,\gamma^{\nu\rho}\}\epsilon \cr&=&{\rm i}
{\bar\epsilon}\{\gamma_{\mu},\, -4\beta e^{\fft32H-\fft12\tilde
H}\Hat\gamma\sigma_{\hat 3}+2 e^{\fft32H}\partial_{\nu}\tilde
H\gamma^{\nu}\sigma_{\hat 1}\}\epsilon =4e^{\fft32  H
}\partial_{\mu}{\tilde H}f_1\,.
\end{eqnarray}
Thus
\begin{equation}
\partial_\mu f_1=\ft12f_1\partial_{\mu}\tilde H \quad\Rightarrow\quad
f_1= c_1 e^{\fft12\tilde H}\,,\qquad\Phi=-c_1e^{2\tilde H}\,.
\label{10Df1}
\end{equation}

From  (\ref{10dbeq2}) and  (\ref{10dbeq3}), we find
\begin{equation}
\gamma^{\mu}\partial_\mu(H+\tilde H)\epsilon=-2i(\alpha e^{-\fft{
H}2}\hat\gamma\sigma_{\hat 1}+\beta e^{-\fft{\tilde
H}2}\hat\gamma\sigma_{\hat 2})\epsilon \label{10ddHH1}
\end{equation}
Thus
\begin{eqnarray}
{c_1}\partial_\mu(H+\tilde H)  e^{\fft12\tilde
H}&=&\partial_\mu(H+\tilde H)f_1=\ft{{\rm i}}2\partial_\nu(H+\tilde
H)\bar\epsilon\{\gamma_{\mu},\gamma^{\nu}\}\sigma_{\hat 1}\epsilon
\cr&=&-{\rm i}\bar\epsilon[\gamma_{\mu},\,{\rm i}\alpha e^{-\fft{
H}2}\hat\gamma- a\beta e^{-\fft{\tilde H}2}]\epsilon =2\alpha
e^{-\fft{ H}2}L_{\mu}\,, \cr c_2\partial_\mu(H+\tilde H)  e^{\fft12
H}&=&\partial_\mu(H+\tilde H)f_2=\ft{{\rm i}}2\partial_\nu(H+\tilde
H)\bar\epsilon \{\gamma_{\mu},\gamma^{\nu}\}\sigma_{\hat 2}\epsilon
\cr&=&-{{\rm i}}\bar\epsilon [\gamma_{\mu},\,a\alpha e^{-\fft{
H}2}+{\rm i}\beta e^{-\fft{\tilde H}2}\hat\gamma]\epsilon =2\beta
e^{-\fft{\tilde  H}2}L_{\mu}\,.
\end{eqnarray}
By changing the overall sign of the 5-form flux and an appropriate
rescaling of the Killing spinor, we can set
\begin{equation}
c_1=\alpha\,.
\end{equation}
It follows that
\begin{equation}
e^{\fft12(H+\tilde H)}=y+ c_3\,,\qquad c_2=\beta\,. \label{10dH+tH}
\end{equation}
By an appropriate shift of $y$, we can set $c_3=0$. Now we find that
\begin{equation}
h^{-2}=-K^2=f_1^2+f_2^2=e^H+e^{\tilde H}\,.
\end{equation}

The equations $K^t=1$ and $L_y =1$ imply that $\epsilon^\dagger
\epsilon =h^{-1}$ and $\epsilon^\dagger \gamma^{\hat0} \hat\gamma
\gamma^{\hat3} \epsilon =h^{-1}$ respectively. Thus we find
\begin{eqnarray}
0=\epsilon^\dagger\left[1-  \gamma^{\hat0} \hat\gamma
\gamma^{\hat3}\right]\epsilon=\ft12\epsilon^\dagger\left[1-
\gamma^{\hat0} \hat\gamma  \gamma^{\hat3}\right]^\dagger\left[1-
\gamma^{\hat0} \hat\gamma  \gamma^{\hat3}\right]\epsilon
\end{eqnarray}
It follows
\begin{eqnarray}
\left[1-  \gamma^{\hat0} \hat\gamma
\gamma^{\hat3}\right]\epsilon=0,\quad\mbox{or}\quad \left[1- {\rm i}
\gamma^{\hat1} \gamma^{\hat2}\right]\epsilon=0 \label{10dprj2}
\end{eqnarray}
Substituting (\ref{10dH+tH}) back to (\ref{10ddHH1}), we find
\begin{eqnarray}
&&\left(\sqrt{1+e^{H-\tilde H}}\gamma^{\hat3}\sigma_{\hat 1}+  {\rm
i}\alpha  \hat\gamma-\lambda\beta e^{\fft{H-\tilde
H}2}\right)\epsilon=0\cr &&\Rightarrow \left( e^{-2 \lambda\beta
\xi\,\gamma^{\hat3}\sigma_{\hat 1}}+ {\rm i}\alpha
\,\gamma^{\hat3}\hat\gamma\sigma_{\hat 1}\right)\epsilon=0
\,.\label{10dprj1}
\end{eqnarray}
where $\sinh(2\xi) =  e^{ \fft12(H-\tilde H)}$. It implies that the
pseudo-Killing spinor has the form
\begin{equation}\label{10Depsilonone}
\epsilon = e^{ \lambda\beta\xi\,\gamma^{\hat3}\sigma_{\hat 1} }
\epsilon_1\,,\qquad (1+ {\rm i}
\alpha\,\gamma^{\hat3}\hat\gamma\sigma_{\hat 1})  \epsilon_1 = 0\,.
\end{equation}
Inserting (\ref{10Depsilonone}) into the expression (\ref{10Df1})
for $f_1$ gives
\begin{equation}
\alpha e^{\fft12\tilde H}={\rm i}\bar\epsilon_1 e^{-\lambda\beta
\gamma^{\hat3} \sigma_{\hat 1} \,\xi }\sigma_{\hat 1}
e^{\lambda\beta \gamma^{\hat3} \sigma_{\hat 1} \,\xi } \epsilon_1 =
{\rm i}\epsilon_1^{\dagger}\gamma^{\hat 0}\sigma_{\hat
1}\epsilon_1\,.
\end{equation}
Thus
\begin{eqnarray} \label{epsilonzero}
&&\epsilon_1 =  e^{\fft14\tilde  H} \epsilon_0\,,\qquad
\epsilon_0^\dagger\gamma^{\hat 0}\sigma_{\hat 1}\epsilon_0 ={\rm
i}\alpha \,,\cr &&\epsilon^t\epsilon= e^{\fft12\tilde  H}
\epsilon_0^t(\cosh2\xi+\sinh2\xi\,\gamma^{\hat 3}\sigma_{\hat 1})
\epsilon_0 \cr&&\,=e^{\fft12\tilde  H}(\cosh2\xi \,\epsilon_0^t
\epsilon_0-{\rm i}\sinh2\xi\,\bar\epsilon_0^c\gamma^{\hat
2}\hat\gamma \sigma_{\hat 1}\epsilon_0) =h^{-1}\,\epsilon_0^t
\epsilon_0\,.
\end{eqnarray}

From the above expressions for Killing spinor we find
\begin{eqnarray}
L^c_{\hat 0} & = & \epsilon^t\gamma_{\hat2} \gamma_{\hat0}\epsilon =
\bar \epsilon^c\hat\gamma \gamma_{\hat3}\sigma_{\hat 1} \epsilon=0
\cr L^c_{\hat 1} & = & \epsilon^t\gamma_{\hat2}
\gamma_{\hat1}\epsilon ={\rm i} \, \epsilon^t\epsilon=  {\rm i}
h^{-1} \epsilon^t_0\epsilon_0 \cr L^c_{\hat 2} &=&  \epsilon^t
\gamma_{\hat2} \gamma_{\hat2} \epsilon= \epsilon^t\epsilon = h^{-1}
\epsilon^t_0\epsilon_0 \cr L^c_{\hat 3} & = &
\epsilon^t\gamma_{\hat2} \gamma_{\hat3}\epsilon = \bar
\epsilon^c\hat\gamma \gamma_{\hat0}\sigma_{\hat 1} \epsilon=0 \cr
L^c & = & L^c_{\hat \nu} e^{\hat \nu}_\mu dx^\mu =  (
\epsilon^t_0\epsilon_0) ({\rm i}{\tilde e}^{\hat 1}_i+  {\tilde
e}^{\hat 2}_i) dx^i
\end{eqnarray}
Where $\tilde e^{\hat c}_i $ is the vielbein of the metric $\tilde
g_{ij} = {\tilde e}^{\hat c}_i {\tilde e}^{\hat c}_{j} $ and
$e^{\hat i}_i = h {\tilde e}^{\hat i}_j$ is the full vielbein for
the four dimensional metric in the directions 1,2. The equation
$dL^c=0$ implies that $ \epsilon_0$ is independent of the time $t$.
Then we can make it as a constant spinor by setting the phase of
$\epsilon_0$ to zero {\it via} a local Lorentz rotation in the
$(x_1,x_2)$-plane. Under this gauge choice, the equation $dL=0$
further implies that the vielbeins $ {\tilde e}^{\hat c}_i$ are
independent of $y$ and that the two dimensional metric is flat. So
we  may choose coordinates such that $\tilde g_{ij} = \delta_{ij}$.

From (\ref{10ddk}) we find that
\begin{eqnarray}
d[h^{-2}(dt+V)]&=&- dK=\ft{1}{2} e^{- \fft32 H }\left(f_1*_4 F+a
f_2\,F \right) \cr&=&\ft{1}{2}\alpha  e^{\fft{\tilde  H-3H}2}
\left[h^2*_3 dB_t+e^{\fft32(H-\tilde H)}(dt+V)\wedge d\Phi\right]\
\cr&& +\ft{1}{2}\lambda \beta e^{- H} \left[d
B_t\wedge(dt+V)+h^{2}e^{\fft32(H-\tilde H)}*_3d\Phi\right] \,.
\end{eqnarray}
It follows that
\begin{eqnarray}
h^{-2}dV&=&\ft{1}{2} h^2 \left[\alpha e^{\fft{\tilde  H-3H}2}*_3
dB_t+\lambda\beta\,e^{\fft{H-3\tilde H}{2}}*_3d\Phi\right]
\cr&=&\lambda\alpha\beta h^2y*_3 d(H-\tilde H)\,.
\end{eqnarray}
Let
\begin{equation}
H_{\pm}=\ft12(H\pm\tilde H)\,,\qquad D=\ft12\lambda\alpha\beta \tanh H_-\,,
\end{equation}
we find
\begin{equation}
dV=2\lambda\alpha\beta\left(e^H+e^{\tilde H}\right)^{-2}y*_3
dH_-=y^{-1}*_3 dD \,.
\end{equation}
Finally, the consistency condition $d^2V=0$ implies
\begin{equation}
\partial^2_iD+y\partial_y\left(\fft1y\partial_yD\right)=0 \,.
\label{d10Deom}
\end{equation}
Thus the solution is completely solved up to (\ref{d10Deom}), which
is the Laplace equation.  This basic equation is identical to that
of the LLM bubbling solution in type IIB supergravity.  We summarize
the solution in (\ref{d10sol1}) and (\ref{10dbeq}).  The $H$ in
(\ref{d10sol1}) should be $H_-$ as the logical consequence of our
derivation, but we remove the subscript in (\ref{d10sol1}) for the
stylistic reason.  We also dropped $\lambda$ since it appears always
together with $\beta$.

\section{Detail derivation of the less-bubbling AdS$_4$ solution}
\label{appB}

In this section, we give the detail derivation of the less-bubbling
solution presented in section 8. As shown in section 6, the critical
dimension for $n=4$ is $D=11$, corresponding to M-theory.  For other
dimensions, additional constraints have to be imposed in order for
the existence of the pseudo-Killing spinors to imply the equations
of motion. We shall consider the case with $D=8$.  The
pseudo-Killing spinor equation is given by
\begin{eqnarray}
0&=&D_M  \eta
+\fft1{96\sqrt3}\left(\Gamma_M{}^{M_1M_2M_3M_4}-4\delta_M^{M_1}
\Gamma^{M_2M_3M_4}\right)F_{M_1M_2M_3M_4}\eta\cr &=&D_M  \eta
+\fft1{96\sqrt3}\Gamma^{M_1M_2M_3M_4}\Gamma_MF_{M_1M_2M_3M_4}\eta\,.
\label{ksdefd8n4}
\end{eqnarray}
The integrability condition can be found in section 6. We begin with
the following ansatz
\begin{eqnarray}
&&ds^2=g_{\mu\nu}dx^{\mu}dx^{\nu}+e^{H}d\Omega_2^2+e^{\tilde H}d\tilde\Omega_2^2\,,
\cr&&
F_{\4}=\ft12 F_{\mu\nu}dx^{\mu}\wedge dx^{\nu}\wedge \Omega_\2
+\ft12{\tilde F}_{\mu\nu}dx^{\mu}\wedge dx^{\nu}\wedge \tilde\Omega_\2
\,,
\end{eqnarray}
which has the $SO(3)\times SO(3)$ isometry. The gamma matrices can
be decomposed as follows
\begin{equation}
\Gamma_{\mu}=\gamma_{\mu}\otimes 1\otimes 1\,,\qquad\Gamma_a=
\hat\gamma \otimes \sigma_a\otimes 1\,,\qquad\Gamma_{\tilde a}=
\hat\gamma \otimes \sigma_{\hat 3}\otimes \sigma_{\tilde a}\,,
\end{equation}
where
\begin{equation}
\hat\gamma ={\rm i}\gamma_{\hat0\hat1\hat2\hat3}=-\fft1{4!}{\rm i}
 \epsilon_{\mu_1\mu_2\mu_3\mu_4}
 \gamma^{\mu_1\mu_2\mu_3\mu_4}\,,
\qquad\epsilon_{\hat0\hat1\hat2\hat3}=1\,,\qquad
\sigma_{\hat 3}={\rm i}\sigma_{\hat1}\sigma_{\hat2}\,.
\end{equation}
where the hatted indices are the velbein indicesand the $\sigma's$
with hatted subscripts are Pauli matrices.  We now perform the
reduction on $S^2\times S^2$ and introduce two component spinors
$\chi_{\pm}$ and $\tilde\chi_{\pm}$ on the two $S^2$, which obey the
equations
\begin{equation}\label{SpinorOnS2}
{\nabla}'_a \chi_\pm=\alpha\frac{{\rm i}}{2}{\sigma}_{\hat
a}\chi_\alpha\,,\qquad {\nabla}'_{\tilde a}
\chi_\beta=\beta\frac{{\rm i}}{2}{\sigma}_{\hat{\tilde
a}}\chi_\beta\,.
\end{equation}
Note that when $\alpha$ and $\beta$ appear as the indices for
$\chi_\alpha$ and $\tilde \chi_\beta$ respectively, they denote
$\pm$; when they appear as numbers, they are $\pm 1$. The
${\nabla}'$ is the covariant derivative defined on a unit 2-sphere.
It is related to the covariant derivative in the sphere directions
of our ansatz as
\begin{equation} \label{spincon}
\nabla_a={\nabla'}_a-\ft{1}{4}{\Gamma^\mu}_a\partial_\mu H,\qquad
\nabla_{\tilde a}={\nabla'}_{\tilde a}-
\ft{1}{4}{\Gamma^\mu}_{\tilde a}\partial_\mu \tilde H,
\end{equation}
The covariance under the $SU(2)$ transformations ensures that we can
take, without loss of generality, that
\begin{equation}
\chi_-={\rm i}\sigma_{\hat 3}\chi_+, \qquad\tilde\chi_-={\rm
i}\sigma_{\hat 3}\tilde\chi_+ \,.
\end{equation}
Now we expand the spinor $\eta$ over basis of the tensor product of
the real Killing spinors on the two $S^2$,
\begin{equation}\label{DecompSpinXi1}
\xi=\sum_{\alpha,\beta=+,-}\epsilon_{\alpha\beta}
\otimes\chi_{\alpha}\otimes\tilde \chi_{\beta} .
\end{equation}
The expression in (\ref{ksdefd8n4}) involving the 4-form becomes
\begin{eqnarray}
&&\fft1{96\sqrt3}\Gamma^{M_1M_2M_3M_4} F_{M_1M_2M_3M_4}
=\fft1{16\sqrt3} ( e^{-  H }
F_{\mu\nu}\Gamma^{\mu\nu}\epsilon_{ab}\Gamma^{ab} +  e^{- \tilde H
}{\tilde F}_{\mu\nu}\Gamma^{\mu\nu} \epsilon_{{\tilde a}{\tilde
b}}\Gamma^{{\tilde a}{\tilde b}}) \cr& =&\fft1{8\sqrt3} ( e^{-  H }
F_{\mu\nu}\gamma^{\mu\nu}\otimes {\rm i}\sigma_{\hat 3}\otimes1 +
e^{- \tilde H }{\tilde F}_{\mu\nu}\gamma^{\mu\nu}\otimes1\otimes
{\rm i}\sigma_{\hat 3})
\end{eqnarray}
The pseudo-Killing spinor equation implies
\begin{eqnarray}
&&\sum_{\alpha,\beta}\left[\nabla_{\rho}\epsilon_{\alpha\beta}
\otimes\chi_{\alpha}\otimes\tilde \chi_{\beta} +\ft1{8\sqrt3} ( e^{-
H }
F_{\mu\nu}\gamma^{\mu\nu}\gamma_{\rho}\epsilon_{\alpha\beta}\otimes
{\rm i}\sigma_{\hat 3}\chi_{\alpha}\otimes\tilde
\chi_{\beta}\right.\cr &&\qquad\qquad\left. +  e^{- \tilde H
}{\tilde
F}_{\mu\nu}\gamma^{\mu\nu}\gamma_{\rho}\epsilon_{\alpha\beta}\otimes
\chi_{\alpha}\otimes {\rm i}\sigma_{\hat 3}\tilde
\chi_{\beta})\right]=0\cr &\Rightarrow&\nabla_{\rho}\epsilon_{++}
+\ft1{8\sqrt3} (- e^{-  H }
F_{\mu\nu}\gamma^{\mu\nu}\gamma_{\rho}\epsilon_{-+} -  e^{- \tilde H
}{\tilde F}_{\mu\nu}\gamma^{\mu\nu}\gamma_{\rho}\epsilon_{+-})=0\,,
\cr&&\nabla_{\rho}\epsilon_{--} +\ft1{8\sqrt3} (e^{-  H }
F_{\mu\nu}\gamma^{\mu\nu}\gamma_{\rho}\epsilon_{+-} +  e^{- \tilde H
}{\tilde F}_{\mu\nu}\gamma^{\mu\nu}\gamma_{\rho}\epsilon_{-+})=0\,,
\cr&&\nabla_{\rho}\epsilon_{+-} +\ft1{8\sqrt3} (- e^{-  H }
F_{\mu\nu}\gamma^{\mu\nu}\gamma_{\rho}\epsilon_{--} +  e^{- \tilde H
}{\tilde F}_{\mu\nu}\gamma^{\mu\nu}\gamma_{\rho}\epsilon_{++})=0\,,
\cr&&\nabla_{\rho}\epsilon_{-+} +\ft1{8\sqrt3} (e^{-  H }
F_{\mu\nu}\gamma^{\mu\nu}\gamma_{\rho}\epsilon_{++} -  e^{- \tilde H
}{\tilde F}_{\mu\nu}\gamma^{\mu\nu}\gamma_{\rho}\epsilon_{--})=0\,,
\end{eqnarray}
and
\begin{eqnarray}
&&\!\!\!\sum_{\alpha,\beta}\left[\ft{{\rm i}\alpha}2e^{-\fft{H}2}
(\epsilon_{\alpha\beta}\otimes \sigma_a\chi_{\alpha}\otimes\tilde
\chi_{\beta}) -\ft14\partial_{\mu}H(\gamma^{\mu} \hat\gamma
\epsilon_{\alpha\beta}\otimes \sigma_a\chi_{\alpha}\otimes\tilde
\chi_{\beta}) \right.\cr&&\!\!\!\!\!\!\left. +\ft1{8\sqrt3} ( e^{- H
} F_{\mu\nu}\gamma^{\mu\nu}\hat\gamma\epsilon_{\alpha\beta}\otimes
{\rm i}\sigma_{\hat 3}\sigma_a\chi_{\alpha}\otimes\tilde
\chi_{\beta} +  e^{- \tilde H }{\tilde
F}_{\mu\nu}\gamma^{\mu\nu}\hat\gamma\epsilon_{\alpha\beta}\otimes\sigma_a
\chi_{\alpha}\otimes {\rm i}\sigma_{\hat 3}\tilde
\chi_{\beta})\right]=0 \cr&\Rightarrow&\!\!\!\!\!\!\left({\rm i}
e^{-\fft{H}2}-\fft12\partial_{\mu}H\gamma^{\mu} \hat\gamma
\right)\epsilon_{++} +\ft1{4\sqrt3} ( e^{-  H }
F_{\mu\nu}\gamma^{\mu\nu}\hat\gamma\epsilon_{-+} -  e^{- \tilde H
}{\tilde F}_{\mu\nu}\gamma^{\mu\nu}\hat\gamma\epsilon_{+-})=0\,,
\cr&&\!\!\!\!\!\!\left(-{\rm
i}e^{-\fft{H}2}-\fft12\partial_{\mu}H\gamma^{\mu} \hat\gamma
\right)\epsilon_{--} +\ft1{4\sqrt3} (-e^{-  H }
F_{\mu\nu}\gamma^{\mu\nu} \hat\gamma \epsilon_{+-} +  e^{- \tilde H
}{\tilde F}_{\mu\nu}\gamma^{\mu\nu} \hat\gamma \epsilon_{-+})=0\,,
\cr&&\!\!\!\!\!\!\left({\rm
i}e^{\fft{H}2}-\fft12\partial_{\mu}H\gamma^{\mu} \hat\gamma
\right)\epsilon_{+-} +\ft1{4\sqrt3} (e^{-  H }
F_{\mu\nu}\gamma^{\mu\nu} \hat\gamma \epsilon_{--}
+  e^{- \tilde H }{\tilde F}_{\mu\nu}\gamma^{\mu\nu} \hat\gamma \epsilon_{++})=0\,,\\
&&\!\!\!\!\!\!\left(-{\rm
i}e^{-\fft{H}2}-\fft12\partial_{\mu}H\gamma^{\mu} \hat\gamma
\right)\epsilon_{-+} +\ft1{4\sqrt3} (-e^{-  H }
F_{\mu\nu}\gamma^{\mu\nu} \hat\gamma \epsilon_{++} -  e^{- \tilde H
}{\tilde F}_{\mu\nu}\gamma^{\mu\nu} \hat\gamma
\epsilon_{--})=0\,,\nonumber
\end{eqnarray}
and
\begin{eqnarray}
&&\!\!\!\!\!\!\sum_{\alpha,\beta}\left[\fft{{\rm
i}\beta}2e^{-\fft{\tilde H}2}(\epsilon_{\alpha\beta}\otimes
\chi_{\alpha}\otimes\sigma_{\tilde a}\tilde \chi_{\beta})
-\ft14\partial_{\mu}{\tilde H}(\gamma^{\mu} \hat\gamma
\epsilon_{\alpha\beta}\otimes \sigma_{\hat
3}\chi_{\alpha}\otimes\sigma_{\tilde a}\tilde \chi_{\beta})
\right.\cr&&\!\!\!\!\!\!\left. +\ft1{8\sqrt3} ({\rm i} e^{-  H }
F_{\mu\nu}\gamma^{\mu\nu}\hat\gamma\epsilon_{\alpha\beta}\otimes
\chi_{\alpha}\otimes\sigma_{\tilde a}\tilde \chi_{\beta} +  e^{-
\tilde H }{\tilde
F}_{\mu\nu}\gamma^{\mu\nu}\hat\gamma\epsilon_{\alpha\beta}
\otimes\sigma_{\hat 3}\chi_{\alpha}\otimes i\sigma_{\hat
3}\sigma_{\tilde a}\tilde \chi_{\beta})\right]=0\,,
\cr&\Rightarrow&\!\!\!\!\!\!\left(e^{-\fft{\tilde
H}2}+\ft1{4\sqrt3}e^{-  H }
F_{\mu\nu}\gamma^{\mu\nu}\hat\gamma\right)\epsilon_{++}
-\ft12\partial_{\mu}{\tilde H}\gamma^{\mu} \hat\gamma \epsilon_{-+}
+\ft1{4\sqrt3} e^{- \tilde H }{\tilde
F}_{\mu\nu}\gamma^{\mu\nu}\hat\gamma\epsilon_{--}=0\,,
\cr&&\!\!\!\!\!\!\left(-e^{-\fft{\tilde H}2}+\ft1{4\sqrt3}e^{-  H }
F_{\mu\nu}\gamma^{\mu\nu}\hat\gamma\right)\epsilon_{--}
+\ft12\partial_{\mu}{\tilde H}\gamma^{\mu} \hat\gamma \epsilon_{+-}
+\ft1{4\sqrt3} e^{- \tilde H }{\tilde
F}_{\mu\nu}\gamma^{\mu\nu}\hat\gamma\epsilon_{++}=0\,,
\cr&&\!\!\!\!\!\!\left(-e^{-\fft{\tilde H}2}+\ft1{4\sqrt3}e^{-  H }
F_{\mu\nu}\gamma^{\mu\nu}\hat\gamma\right)\epsilon_{+-}
-\ft12\partial_{\mu}{\tilde H}\gamma^{\mu} \hat\gamma \epsilon_{--}
-\ft1{4\sqrt3} e^{- \tilde H }{\tilde
F}_{\mu\nu}\gamma^{\mu\nu}\hat\gamma\epsilon_{-+}=0\,,
\cr&&\!\!\!\!\!\!\left(e^{-\fft{\tilde H}2}+\fft1{4\sqrt3}e^{-  H }
F_{\mu\nu}\gamma^{\mu\nu}\hat\gamma\right)\epsilon_{-+}
+\ft12\partial_{\mu}{\tilde H}\gamma^{\mu} \hat\gamma \epsilon_{++}
-\ft1{4\sqrt3} e^{- \tilde H }{\tilde
F}_{\mu\nu}\gamma^{\mu\nu}\hat\gamma\epsilon_{+-}=0\,.
\end{eqnarray}
Great simplification occurs if we take $\tilde F_{(2)}=0$, in which
case, $\{\epsilon_{++},\epsilon_{-+}\}$ decouple from
$\{\epsilon_{--},\epsilon_{+-}\}$.  We can rewrite the above
equations as
\begin{eqnarray}
&&\nabla_{\rho}
\begin{pmatrix}
\epsilon_{++} \cr
 \hat\gamma \epsilon_{-+}
\end{pmatrix}
+\fft1{8\sqrt3} e^{-  H }  F_{\mu\nu} \hat\gamma \gamma^{\mu\nu}\gamma_{\rho}
\begin{pmatrix}
 \hat\gamma \epsilon_{-+}\cr
\epsilon_{++}
\end{pmatrix}=0\,,
\\&&\left({\rm i}e^{-\fft{H}2}-\fft12\partial_{\mu}H\gamma^{\mu}
 \hat\gamma \right)
\begin{pmatrix}
\epsilon_{++} \cr
 \hat\gamma \epsilon_{-+}
\end{pmatrix}
+\fft1{4\sqrt3}  e^{-  H }  F_{\mu\nu}\gamma^{\mu\nu}
\begin{pmatrix}
 \hat\gamma \epsilon_{-+}\cr
\epsilon_{++}
\end{pmatrix}=0\,,
\\&&
\left(e^{-\fft{\tilde H}2}+\fft1{4\sqrt3}e^{-  H }
F_{\mu\nu}\gamma^{\mu\nu}\hat\gamma\right)
\begin{pmatrix}
\epsilon_{++} \cr
 \hat\gamma \epsilon_{-+}
\end{pmatrix}
-\fft12\partial_{\mu}{\tilde H}\gamma^{\mu}
\begin{pmatrix}
 \hat\gamma \epsilon_{-+}\cr
\epsilon_{++}
\end{pmatrix}=0\,,
\end{eqnarray}
and
\begin{eqnarray}
&&\nabla_{\rho}
\begin{pmatrix}
\epsilon_{+-} \cr
 \hat\gamma \epsilon_{--}
\end{pmatrix}
+\fft1{8\sqrt3} e^{-  H }  F_{\mu\nu} \hat\gamma \gamma^{\mu\nu}\gamma_{\rho}
\begin{pmatrix}
 \hat\gamma \epsilon_{--}\cr
\epsilon_{+-}
\end{pmatrix}=0\,,
\\&&\left({\rm i}e^{-\fft{H}2}-\fft12\partial_{\mu}H\gamma^{\mu}
\hat\gamma \right)
\begin{pmatrix}
\epsilon_{+-} \cr
 \hat\gamma \epsilon_{--}
\end{pmatrix}
+\fft1{4\sqrt3}  e^{-  H }  F_{\mu\nu}\gamma^{\mu\nu}
\begin{pmatrix}
 \hat\gamma \epsilon_{--}\cr
\epsilon_{+-}
\end{pmatrix}=0\,,
\\&&
\left(-e^{-\fft{\tilde H}2}+\fft1{4\sqrt3}e^{-  H }
F_{\mu\nu}\gamma^{\mu\nu}\hat\gamma\right)
\begin{pmatrix}
\epsilon_{+-} \cr
 \hat\gamma \epsilon_{--}
\end{pmatrix}
-\fft12\partial_{\mu}{\tilde H}\gamma^{\mu}
\begin{pmatrix}
 \hat\gamma \epsilon_{--}\cr
\epsilon_{+-}
\end{pmatrix}=0\,.
\end{eqnarray}
Now it is easy to see that there are in fact four independent combinations
\begin{equation}
\epsilon_{++}\pm \hat\gamma \epsilon_{-+} \,,\qquad\epsilon_{+-}\pm
\hat\gamma \epsilon_{--}\,.
\end{equation}
Any one but only one of the above four independent combinations can
be the pseudo-Killing spinor. We can thus choose one combination and
drop the subscript, giving rise to the following equations.
\begin{eqnarray}
&&\left[\nabla_{\rho}
+\fft1{8\sqrt3} e^{-  H }  F_{\mu\nu} \hat\gamma \gamma^{\mu\nu}\gamma_{\rho}\right]
\epsilon=0\,,\label{8Dbeq1}
\\&&\left[{\rm i}e^{-\fft{H}2}-\ft12\partial_{\mu}H\gamma^{\mu} \hat\gamma
+\fft1{4\sqrt3}  e^{-  H }  F_{\mu\nu}\gamma^{\mu\nu}\right]
\epsilon=0\,,\label{8Dbeq2}
\\&&
\left[me^{-\fft{\tilde H}2}+\fft1{4\sqrt3}e^{-  H }
F_{\mu\nu}\gamma^{\mu\nu}\hat\gamma -\ft12\partial_{\mu}{\tilde
H}\gamma^{\mu}\right] \epsilon=0\,,\label{8Dbeq3}
\end{eqnarray}
where $m=\pm1$.

We are now in the position to proceed to derive the bubbling
solution. As in the previous example in appendix A, we start by
defining the following real spinor bilinears
\begin{equation}
f_1={\rm i}{\bar\epsilon}\epsilon,\quad f_2={\bar\epsilon}
\hat\gamma \epsilon,\quad
K_\mu={\bar\epsilon}\gamma_\mu\epsilon,\quad
L_\mu={\bar\epsilon}\gamma_\mu \hat\gamma \epsilon, \quad
Y_{\mu\nu}={\bar\epsilon}\gamma_{\mu\nu}\epsilon,
\end{equation}
as well as the complex spinor bilinears
\begin{eqnarray}
&& L^c_\mu={\bar\epsilon}^c\gamma_\mu\epsilon,
\quad Y^c_{\mu\nu}={\bar\epsilon}^c\gamma_{\mu\nu}\epsilon,
\end{eqnarray}
where $\bar\epsilon=\epsilon^+\gamma^{\hat 0}$ and
$\bar\epsilon^c=\epsilon^t \gamma_{\hat2}$. We find
\begin{eqnarray}
\nabla_{\mu}f_1 &=&-\fft{1}{4\sqrt3} e^{-  H
}\epsilon_{\mu}{}^{\nu\rho\sigma} F_{\nu\rho}K_{\sigma}\,,
\label{8ddf1}
\\
\nabla_{\mu}f_2
&=&\fft{1}{2\sqrt3} e^{-  H }F_{\mu\nu} K^{\nu}\,, \label{8ddf2}
\\
\nabla_{\mu}K_{\nu}&=&\fft{1}{4\sqrt3} e^{-  H }
\left(\epsilon_{\mu\nu}{}^{\lambda\rho}F_{\lambda\rho}f_1
-2F_{\mu\nu} f_2\right)\,,\label{8Ddk}
\\
\nabla_{\mu}L_{\nu}&=&-\fft{1}{4\sqrt3} e^{-  H }
\left(g_{\mu\nu}F_{\lambda\rho}Y^{\lambda\rho}
+2F_{\mu\rho}Y^{\rho}{}_{\nu}+2F_{\nu\rho}Y^{\rho}{}_{\mu}\right)\,,
\\
\nabla_{\mu}L^c_{\nu}&=&\fft{1}{4\sqrt3} e^{-  H }
\left(g_{\mu\nu}F^{\lambda\rho}Y^c_{\lambda\rho}
+2F_{\mu}{}^{\rho}Y^{c}_{\rho\nu}+2F_{\nu}{}^{\rho}Y^{c}_{\rho\mu}\right)\,.
\end{eqnarray}
Thus $K^{\mu}$ is a Killing vector while the form
$L=L_{\mu}dx^{\mu}$ and $L^c=L^c_{\mu}dx^{\mu}$ are (locally) exact.
Using Fierz identities we have
\begin{equation}
K\cdot L=0\,,\qquad
L^2=-K^2=f_1^2+f_2^2\,.
\end{equation}

Since $L=L_{\mu}dx^{\mu}$ is (locally) exact, we can choose a
coordinate $y$ through
\begin{equation}
  dy=L_\mu dx^\mu
\end{equation}
The other three coordinates can be chosen to lie in the subspace
orthogonal to $y$ such that
\begin{equation}
ds^2=h^2 dy^2+g_{\alpha\beta}dx^\alpha dx^\beta\,,\qquad h^{-2}=L^2=f_1^2+f_2^2\,.
\end{equation}
Let us now look at the Killing vector $K^\mu$. Using the relation
\begin{equation}
0=K^\mu L_\mu =  K^y L_y=  K^y
\end{equation}
we find that $K^\alpha$ is a vector in the three-dimensional space
spanned by $x^\alpha$. Choosing one of the coordinates along
$K^\alpha$ (we shall call it $t$), the most general metric of the
four-dimensional subspace is given by
\begin{equation}
ds^2=-h^{-2}(dt+V_i dx^i)^2+h^2( dy^2+\tilde g_{ij}dx^i dx^j)
\end{equation}
where $i,j=1,2$. The signature is thus determined by the fact that
the Killing vector $K$ is time-like.

The equation of motion for the 4-form, $d *_{11}G_{\4}=0$, implies that
\begin{equation}\label{8dFldStrEqn}
d(e^{\tilde H-H}*_4 dB)=0
\end{equation}
Let us split the vector field $B$ as
\begin{equation}
B_\mu dx^\mu=B_t dt+B_\alpha dx^\alpha=B_t(dt+V_idx^i)+(B_\alpha-B_t
V_\alpha) dx^\alpha\equiv B_t(dt+V_idx^i)+{\hat B}\,.
\end{equation}
Then its field strength $F=dB$ is
\begin{eqnarray}
F&=&dB_t\wedge(dt+V)+(d{\hat B}+B_t dV)\,,\cr
*_4 F&=& h^{2}*_3 dB_t+h^{-2} (dt+V)\wedge*_3(d{\hat B}+B_t dV)\,,
\end{eqnarray}
where $*_3$ is defined in $ds_3^2=dy^2+\tilde g_{ij}dx^i dx^j$.
Then (\ref{8dFldStrEqn}) implies
\begin{equation}
d*_3\left[h^{-2}e^{\tilde H-H}(d{\hat B}+B_t dV)\right]=0\,.
\end{equation}
This means that locally we can introduce a dual potential $\Phi$:
\begin{equation}\label{8DDefPhiM}
d{\hat B}+B_t dV=h^2\,e^{H-\tilde H}*_3d\Phi
\end{equation}
Thus the vector field $B$ can be described by its time component
$B_t$ together with the potential $\Phi$. The time component of
(\ref{8dFldStrEqn}) leads to the equation:
\begin{equation}
d\left[V\wedge d\Phi+ h^2e^{\tilde H-H}*_3 dB_t\right]=0
\end{equation}

As shown in section 6 and remarked in section 8, there is an
additional constrain (\ref{d8constr}) from the integrability
condition.  For our ansatz, it implies that $F\wedge F=0$, which
gives rise to
\begin{eqnarray}
dB_t\wedge*_3d\Phi=0~~\Rightarrow~~\partial_{\mu}B_t \,\partial^{\mu}\Phi=0\,.
\end{eqnarray}

From the  equation (\ref{8ddf2}) and the fact that $B_i$ is
independent of $t$, we find
\begin{equation}
\partial_\mu f_2=\fft{1}{2\sqrt3}e^{-H}\partial_\mu B_t, \quad \mbox{\it i.e.}\quad
df_2=\fft{1}{2\sqrt3}e^{- H}d B_t\,.
\end{equation}
By using (\ref{8Dbeq2}),  we find
\begin{equation}
\partial_\mu B_t=F_{\mu\nu} K^\nu=
 \ft{1}{4} \bar \epsilon [\gamma_\mu,{\not F}] \epsilon
=\sqrt3 \bar \epsilon [\gamma_\mu,-{\rm
i}e^{\fft{H}2}+\ft12e^H\partial_{\nu}H\gamma^{\nu} \hat\gamma ]
\epsilon ={\sqrt3}(\partial_{\mu} e^H)f_2\,.
\end{equation}
Thus
\begin{equation}
\partial_\mu f_2=\ft12f_2\partial_{\mu}H~~ \Rightarrow~~ f_2
= c_2 e^{\fft12H}\,,\qquad B_t=\fft{2}{\sqrt 3}c_2e^{\fft32 H}\,.
\end{equation}

From the  equation (\ref{8ddf1}), we find
\begin{equation}
\partial_\mu f_1=-\fft{1}{4\sqrt3} e^{-  H }\epsilon_{\mu}{}^{\nu\rho\sigma}
F_{\nu\rho}K_{\sigma}
=-\fft{1}{2\sqrt3}(*_4F)_{\mu\nu}K^{\nu}
=\fft{1}{2\sqrt3}e^{-\tilde H} \partial_{\mu}\Phi\,.
\end{equation}
By using (\ref{8Dbeq3}), we have
\begin{equation}
\epsilon_{\mu\nu\rho\sigma}F^{\nu\rho}K^\sigma=
\fft{{\rm i}}{2}   F_{\nu\rho}
{\bar\epsilon} \hat\gamma \{\gamma_{\mu},\gamma^{\nu\rho}\}\epsilon
=-2{\sqrt3}e^{  H }\partial_{\mu}{\tilde H}f_1
\end{equation}
Thus
\begin{equation}
\partial_\mu f_1=\ft12f_1\partial_{\mu}\tilde H ~~\Rightarrow~~
f_1= c_1 e^{\fft12\tilde H}\,,
\qquad\Phi=\fft{2}{\sqrt 3}c_1e^{\fft32\tilde H}\,.
\label{8Df1}
\end{equation}
By choosing the overall sign of the five-form flux and an
appropriate rescaling of the Killing spinor, we may set
\begin{equation}
c_1=-1\,.
\end{equation}

From (\ref{8Dbeq2}) and  (\ref{8Dbeq3})
\begin{equation}
\gamma^{\mu}\partial_\mu(H+\tilde H)\epsilon=(2m e^{-\fft{\tilde
H}2}-2 i \hat\gamma e^{-\fft{ H}2})\epsilon\,.\label{8DdHH}
\end{equation}
Thus
\begin{eqnarray}
-\partial_\mu(H+\tilde H)  e^{\fft12\tilde
H}&=&\partial_\mu(H+\tilde H)f_1=\fft{{\rm i}}2\partial_\nu(H+\tilde
H)\bar\epsilon\{\gamma_{\mu},\gamma^{\nu}\}\epsilon \cr&=&-\fft{{\rm
i}}2\bar\epsilon[\gamma_{\mu},2m e^{-\fft{\tilde H}2}-2 {\rm i}
\hat\gamma e^{-\fft{ H}2}]\epsilon =-2e^{-\fft{ H}2}L_{\mu}\,, \cr
c_2\partial_\mu(H+\tilde H)  e^{\fft12  H}&=&\partial_\mu(H+\tilde
H)f_2=\ft12\partial_\nu(H+\tilde H)\bar\epsilon \hat\gamma
\{\gamma_{\mu},\gamma^{\nu}\}\epsilon \cr&=&\ft12\bar\epsilon
\hat\gamma \{\gamma_{\mu},2m e^{-\fft{\tilde H}2}-2 {\rm i}
\hat\gamma e^{-\fft{ H}2}\}\epsilon =-2me^{-\fft{\tilde
H}2}L_{\mu}\,.
\end{eqnarray}
It follows that
\begin{equation}
e^{\fft12(H+\tilde H)}=y+ c_3\,,\qquad c_2=-m\,. \label{8DH+tH}
\end{equation}
By an appropriate shift of $y$, we can set $c_3=0$. Now we find that
\begin{equation}
h^{-2}=f_1^2+f_2^2=e^H+e^{\tilde H}\,.
\end{equation}

The equations $K^t=1$ and $L_y =1$ imply that $\epsilon^\dagger
\epsilon =h^{-1}$ and $\epsilon^\dagger \gamma^{\hat0} \hat\gamma
\gamma^{\hat3} \epsilon =h^{-1}$ respectively. Thus we find
\begin{equation}
0=\epsilon^\dagger\left[1-  \gamma^{\hat0} \hat\gamma
\gamma^{\hat3}\right]\epsilon=\ft12\epsilon^\dagger\left[1-
\gamma^{\hat0} \hat\gamma  \gamma^{\hat3}\right]^\dagger\left[1-
\gamma^{\hat0} \hat\gamma  \gamma^{\hat3}\right]\epsilon
\end{equation}
It follows
\begin{equation}
\left[1-  \gamma^{\hat0} \hat\gamma
\gamma^{\hat3}\right]\epsilon=0,\quad\mbox{or}\quad \left[1- {\rm i}
\gamma^{\hat1} \gamma^{\hat2}\right]\epsilon=0 \label{8Dprj2}
\end{equation}
Substituting (\ref{8DH+tH}) back to (\ref{8DdHH}), we find
\begin{eqnarray}
&&\left(\sqrt{1+e^{H-\tilde H}}\gamma^{\hat3}+ {\rm i}
\hat\gamma-me^{\fft{H-\tilde H}2}\right)\epsilon=0\cr &&\Rightarrow
\left( e^{-2 m \gamma^{\hat3}\,\xi }+ {\rm i}
\,\gamma^{\hat3}\hat\gamma\right)\epsilon=0 \,.\label{8Dprj1}
\end{eqnarray}
where $\sinh  2\xi =  e^{ \fft12(H-\tilde H)}$. It implies that the
Killing spinor has the form
\begin{equation}\label{8Depsilonone1}
\epsilon = e^{ m\gamma^{\hat3} \,\xi } \epsilon_1 ~,\qquad
(1+ {\rm i} \,\gamma^{\hat3}\hat\gamma)  \epsilon_1 = 0,\qquad
\end{equation}
Inserting (\ref{8Depsilonone1}) into the expression (\ref{8Df1})
for $f_1$ gives
\begin{equation}
e^{\fft12\tilde H}=i \bar\epsilon_1 e^{-m \gamma^{\hat3}  \,\xi }
e^{m \gamma^{\hat3}  \,\xi } \epsilon_1 =
i\epsilon_1^{\dagger}\gamma^{\hat 0}\epsilon_1\,.
\end{equation}
Thus
\begin{eqnarray} \label{epsilonzero1}
&&\epsilon_1 =  e^{\fft14\tilde  H} \epsilon_0 ~,~~~
\epsilon_0^\dagger\gamma^{\hat 0}\epsilon_0 =-{\rm i} ~,\cr&&
\epsilon^t\epsilon=e^{\fft12\tilde  H}
\epsilon_0^t(\cosh2\xi+\sinh2\xi\,\gamma^{\hat 3}) \epsilon_0
\cr&&\,=e^{\fft12\tilde  H}(\cosh2\xi \,\epsilon_0^t \epsilon_0-{\rm
i}\sinh2\xi\,\bar\epsilon_0^c\gamma^{\hat 2}\hat\gamma \epsilon_0)
=h^{-1}\,\epsilon_0^t \epsilon_0
\end{eqnarray}

From the above expressions for Killing spinor we find
\begin{eqnarray}
L^c_{\hat 0} & = & \epsilon^t\gamma_{\hat2} \gamma_{\hat0}\epsilon =
\bar \epsilon^c\hat\gamma \gamma_{\hat3}\epsilon=0 \cr L^c_{\hat 1}
& = & \epsilon^t\gamma_{\hat2} \gamma_{\hat1}\epsilon ={\rm i} \,
\epsilon^t\epsilon=  {\rm i}   h^{-1} \epsilon^t_0\epsilon_0 \cr
L^c_{\hat 2} &=&  \epsilon^t \gamma_{\hat2} \gamma_{\hat2} \epsilon=
\epsilon^t\epsilon = h^{-1} \epsilon^t_0\epsilon_0 \cr L^c_{\hat 3}
& = & \epsilon^t\gamma_{\hat2} \gamma_{\hat3}\epsilon = \bar
\epsilon^c\hat\gamma \gamma_{\hat0}\epsilon=0 \cr L^c & = &
L^c_{\hat \nu} e^{\hat \nu}_\mu dx^\mu =  ( \epsilon^t_0\epsilon_0)
({\rm i}{\tilde e}^{\hat 1}_i+  {\tilde e}^{\hat 2}_i) dx^i
\end{eqnarray}
Where $\tilde e^{\hat c}_i $ is the vielbein of the metric $\tilde
g_{ij} = {\tilde e}^{\hat c}_i {\tilde e}^{\hat c}_{j} $ and
$e^{\hat i}_i = h {\tilde e}^{\hat i}_j$ is the full vielbein for
the four dimensional metric in the directions 1,2 . The equation
$dL^c=0$ implies that $ \epsilon_0$ is independent of the time $t$.
Then we can make it to be a constant spinor by setting the phase of
$\epsilon_0$ to zero {\it via} a local Lorentz rotation in the
$(x_1,x_2)$-plane. Under this gauge choice, the equation $dL=0$
further implies that the vielbeins $ {\tilde e}^{\hat c}_i$ are
independent of $y$ and that the two dimensional metric is flat. So
we  choose coordinates such that $\tilde g_{ij} = \delta_{ij}$.

From (\ref{8Ddk}) we find that
\begin{eqnarray}
d[h^{-2}(dt+V)]&=&- dK=-\fft{1}{\sqrt3} e^{-  H }\left(f_1*_4
F-f_2\,F \right) \cr&=&\fft{1}{\sqrt3}  e^{\fft{\tilde  H}2-H }
\left[h^2*_3 dB_t+e^{H-\tilde H}(dt+V)\wedge d\Phi\right]\ \cr&&
-\fft{m}{\sqrt3} e^{- \fft{H}2 } \left[d
B_t\wedge(dt+V)+h^{2}e^{H-\tilde H}*_3d\Phi\right] \,.
\end{eqnarray}
It follows that
\begin{eqnarray}
h^{-2}dV&=&\fft{1}{\sqrt3} h^2 \left[e^{\fft{1}2\tilde H-H }*_3
dB_t-m\,e^{\fft12H-\tilde H}*_3d\Phi\right] \cr&=&-m h^2y*_3
d(H-\tilde H)
\end{eqnarray}
Let
\begin{equation}
H_{\pm}=\ft12(H\pm\tilde H)\,,\qquad D=-\ft12m \tanh H_-
\end{equation}
we find
\begin{equation}
dV=-2m\left(e^H+e^{\tilde H}\right)^{-2}y*_3 dH_-=y^{-1}*_3 dD \,.
\end{equation}
Finally, the consistency condition $d^2V=0$ implies
\begin{equation}
\partial^2_iD+y\partial_y\left(\fft1y\partial_yD\right)=0 \,.
\end{equation}
The additional condition is
\begin{equation}
\partial_{\mu}B_t \,\partial^{\mu}\Phi=0\Rightarrow \partial_{\mu}H
\,\partial^{\mu}\tilde H=0
 \Rightarrow
(\partial_{i}D)^2+(\partial_y D)^2=y^{-2}(1-D^2)^2
\end{equation}
We can now assemble the final solution, and we present the result in
section 8. Note that we remove the subscript in the notation $H_-$
in the presentation of the solution. The discussion of the
properties of the solution can be found there.


\begin{thebibliography}{99}
\bibitem{mald}
  J.M.~Maldacena,
{\it The large ${\cal N}$ limit of superconformal field theories and supergravity,}
  Adv.\ Theor.\ Math.\ Phys.\  {\bf 2}, 231 (1998)
  [Int.\ J.\ Theor.\ Phys.\  {\bf 38}, 1113 (1999)]
  [arXiv:hep-th/9711200].

\bibitem{gkp}
  S.S.~Gubser, I.R.~Klebanov and A.M.~Polyakov,
{\it Gauge theory correlators from non-critical string theory,}
  Phys.\ Lett.\  B {\bf 428}, 105 (1998)
  [arXiv:hep-th/9802109].

\bibitem{wit}
  E.~Witten,
{\it Anti-de Sitter space and holography,}
  Adv.\ Theor.\ Math.\ Phys.\  {\bf 2}, 253 (1998)
  [arXiv:hep-th/9802150].

\bibitem{gghpr}
  J.P.~Gauntlett, J.B.~Gutowski, C.M.~Hull, S.~Pakis and H.S.~Reall,
{\it All supersymmetric solutions of minimal supergravity in five dimensions,}
  Class.\ Quant.\ Grav.\  {\bf 20}, 4587 (2003)
  [arXiv:hep-th/0209114].

\bibitem{llm} H. Lin, O. Lunin and J.M. Maldacena,
{\it Bubbling AdS space and $\fft12$-BPS geometries,}
  JHEP {\bf 0410}, 025 (2004)
  [arXiv:hep-th/0409174].

\bibitem{clv}
  S.~Cucu, H.~L\"u and J.F.~Vazquez-Poritz,
{\it Interpolating from AdS$_{D-2} \times S^2$ to AdS$_D$,}
  Nucl.\ Phys.\  B {\bf 677}, 181 (2004)
  [arXiv:hep-th/0304022].

\bibitem{lpv}
  H.~L\"u, C.N.~Pope and J.F.~Vazquez-Poritz,
{\it From AdS black holes to supersymmetric flux-branes,}
  Nucl.\ Phys.\  B {\bf 709}, 47 (2005)
  [arXiv:hep-th/0307001].

\bibitem{psrv}
  J.~Perz, P.~Smyth, T.~Van Riet and B.~Vercnocke,
{\it First-order flow equations for extremal and non-extremal black holes,}
  JHEP {\bf 0903}, 150 (2009)
  [arXiv:0810.1528 [hep-th]].

\bibitem{sketow}
  K.~Skenderis and P.~K.~Townsend,
{\it Pseudo-supersymmetry and the domain-wall/cos- mology correspondence,}
  J.\ Phys.\ A  {\bf 40}, 6733 (2007)
  [arXiv:hep-th/0610253].

\bibitem{ds1}J.~Grover, J.B.~Gutowski, C.A.R.~Herdeiro and W.~Sabra,
{\it HKT Geometry and de Sitter Supergravity,}
  Nucl.\ Phys.\  B {\bf 809}, 406 (2009)
  [arXiv:0806.2626 [hep-th]].

\bibitem{ds2}
  J.~Grover, J.B.~Gutowski, C.A.R.~Herdeiro, P.~Meessen, A.~Palomo-Lozano and W.A.~Sabra,
{\it Gauduchon-Tod structures, Sim holonomy and De Sitter
supergravity,} JHEP {\bf 0907}, 069 (2009)
  [arXiv:0905.3047 [hep-th]].

\bibitem{lpr}
H.~L\"u, C.N.~Pope and J.~Rahmfeld,
{\it A construction of Killing spinors on $S^n$,}
J.\ Math.\ Phys.\  {\bf 40}, 4518 (1999) [arXiv:hep-th/9805151].

\bibitem{lpt} H.~L\"u, C.N.~Pope and P.K.~Townsend,
{\it Domain walls from anti-de Sitter spacetime,}
  Phys.\ Lett.\  B {\bf 391}, 39 (1997)
  [arXiv:hep-th/9607164].

\bibitem{stainless} H.~L\"u, C.N.~Pope, E.~Sezgin and K.S.~Stelle,
{\it Stainless super p-branes,} Nucl.\ Phys.\  B {\bf 456}, 669 (1995)
[arXiv:hep-th/9508042].

\bibitem{guev}
  R.~Gueven,
{\it Black p-brane solutions of $D = 11$ supergravity theory,}
  Phys.\ Lett.\  B {\bf 276}, 49 (1992).

\bibitem{pt}
  G.~Papadopoulos and P.K.~Townsend,
{\it Intersecting M-branes,}
  Phys.\ Lett.\  B {\bf 380}, 273 (1996)
  [arXiv:hep-th/9603087].

\bibitem{tsey}
  A.A.~Tseytlin, {\it Harmonic superpositions of M-branes,}
  Nucl.\ Phys.\  B {\bf 475}, 149 (1996)
  [arXiv:hep-th/9604035].

\bibitem{ggpr}
  U.~Gran, J.~Gutowski, G.~Papadopoulos and D.~Roest,
{\it Systematics of IIB spinorial geometry,}
  Class.\ Quant.\ Grav.\  {\bf 23}, 1617 (2006)
  [arXiv:hep-th/0507087].

\bibitem{gmsw}
  J.P.~Gauntlett, D.~Martelli, J.~Sparks and D.~Waldram,
{\it Supersymmetric AdS$_5$ solutions of type IIB supergravity,}
  Class.\ Quant.\ Grav.\  {\bf 23}, 4693 (2006)
  [arXiv:hep-th/0510125].

\bibitem{fp}
  J.M.~Figueroa-O'Farrill and G.~Papadopoulos,
{\it Maximally supersymmetric solutions of ten-dimensional and
  eleven-dimensional supergravities,}
  JHEP {\bf 0303}, 048 (2003)
  [arXiv:hep-th/0211089].

\bibitem{GP} J.P.~Gauntlett and S.~Pakis,
{\it The Geometry of $D = 11$ Killing spinors,}
  JHEP.\ {\bf 0304}, 039 (2003)
  [arXiv:hep-th/0212008].

\bibitem{GMR} J.B.~Gutowski, D.~Martelli and H.S.~Reall,
{\it All Supersymmetric solutions of minimal supergravity in six- dimensions,}
  Class.\ Quant.\ Grav.\ {\bf 20} 5049 (2003)
  [arXiv:hep-th/0306235]

\bibitem{lv} J.T.~Liu and D.~Vaman,
{\it Bubbling $\fft12$-BPS solutions of minimal six-dimensional supergravity,}
  Phys.\ Lett.\  B {\bf 642}, 411 (2006)
  [arXiv:hep-th/0412242].

\bibitem{ber}
  D.~Berenstein,
{\it A toy model for the AdS/CFT correspondence,}
  JHEP {\bf 0407}, 018 (2004)
  [arXiv:hep-th/0403110].

\bibitem{cjr}
  S.~Corley, A.~Jevicki and S.~Ramgoolam,
{\it Exact correlators of giant gravitons from dual ${\cal N} = 4$
SYM theory,}  Adv.\ Theor.\ Math.\ Phys.\  {\bf 5}, 809 (2002)
  [arXiv:hep-th/0111222].

\bibitem{ccdlllvw}
B.~Chen, S.~Cremonini, A.~Donos, F.-L. Lin, H.~Lin, James T. Liu,
D.~Vaman, Wen-Yu Wen, {\it Bubbling AdS and droplet descriptions of
BPS geometries in IIB supergravity,}
  JHEP {\bf 0710}, 003 (2007)
  [arXiv:0704.2233 [hep-th]].



\end{thebibliography}
\end{document}